\documentclass[a4paper,oneside,12pt]{book}

\newcommand{\thesistitle}{Privacy-Preserving Credit Card Fraud Detection using Homomorphic Encryption} 
\newcommand{\degree}{BA(Mod) in Computer Science} 
\newcommand{\typeofthesis}{report} 
\newcommand{\authorname}{David Nugent} 

\newcommand{\keywords}{homomorphic encryption, credit card fraud, fraud detection, machine learning} 
\newcommand{\school}{\href{http://www.scss.tcd.ie}{School of Computer Science and Statistics}} 

\AtBeginDocument{
\hypersetup{pdftitle=\thesistitle} 
\hypersetup{pdfauthor=\authorname} 
\hypersetup{pdfkeywords=\keywords} 
\hypersetup{pdfsubject=\degree} 
}

\usepackage[T1]{fontenc} 
\usepackage[utf8]{inputenc}
\usepackage[english]{babel}

\usepackage[round,sort,comma,numbers]{natbib}

\usepackage[a4paper,top=2.54cm,bottom=2.54cm,left=2.54cm,right=2.54cm,headheight=16pt]{geometry}
\usepackage{amsmath}
\usepackage[autostyle=true]{csquotes} 
\usepackage[pdftex]{graphicx}
\usepackage[colorinlistoftodos]{todonotes}
\usepackage[colorlinks=true, allcolors=black]{hyperref}
\usepackage{caption} 
\usepackage{sfmath} 
\usepackage[parfill]{parskip}    
\usepackage{setspace} 
\usepackage{enumerate} 
\usepackage{booktabs} 
\usepackage{fancyhdr}
\usepackage{amsthm}
\usepackage{changepage,threeparttable} 
\usepackage[figuresleft]{rotating} 
\usepackage{algorithm} 
\makeatletter
\renewcommand{\ALG@name}{Protocol}
\makeatother
\usepackage{algorithmic}

\setcitestyle{square}

\usepackage{titlesec}
\titleformat{\chapter}[hang] 
{\normalfont\huge\bfseries}{\thechapter}{1cm}{} 

\title{\thesistitle}
\author{\authorname}

\frontmatter
\begin{document}
\begin{titlepage}

\center 



\includegraphics{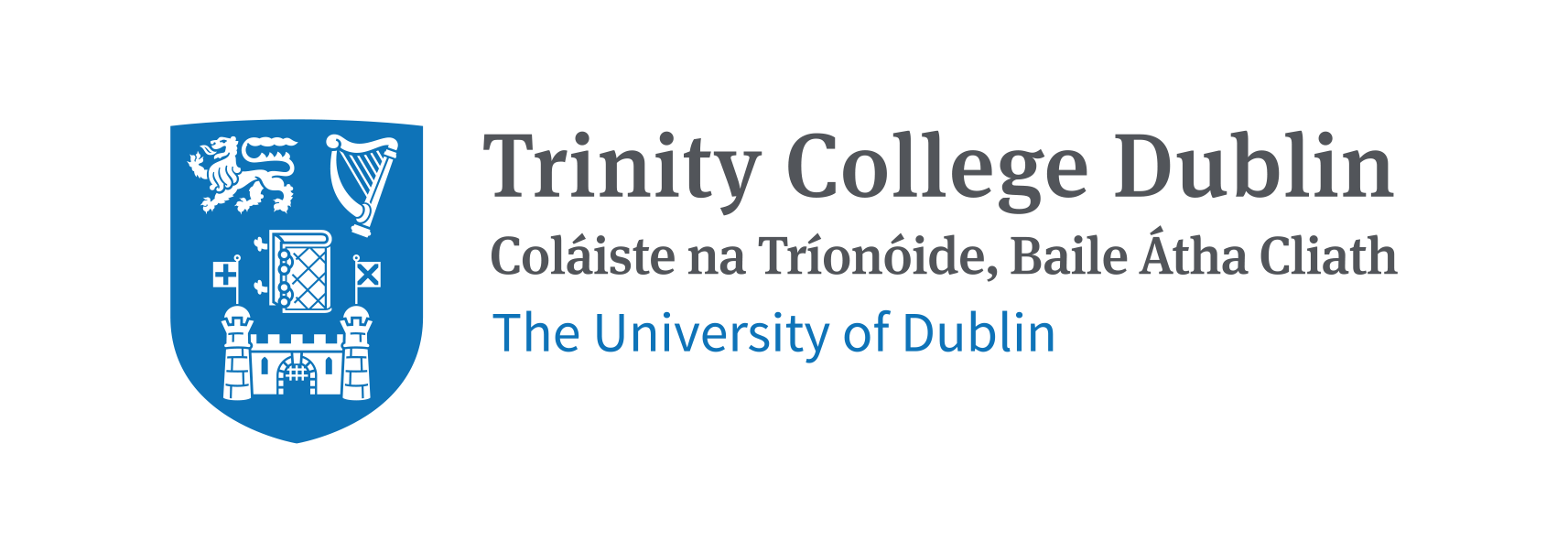}\\[1cm] 

\Large \school\\[1.5cm] 
\ifdefined\department
\large \department\\[1.5cm] 
\fi

\makeatletter
{ \huge \bfseries \thesistitle}\\[1.5cm] 


\ifdefined\authorid
\authorname\\ 
\authorid\\[2cm] 
\else
\authorname\\[2cm] 
\fi


{\large \today}\\[2cm] 

\vfill
 A \typeofthesis\ submitted in partial fulfilment\\of the requirements for the degree of\\
\degree

\vfill 

\end{titlepage}
\pagenumbering{roman}




\chapter*{Abstract}

Credit card fraud is a problem continuously faced by financial institutions and their customers, which is mitigated by fraud detection systems. However, these systems require the use of sensitive customer transaction data, which introduces both a lack of privacy for the customer and a data breach vulnerability to the card provider. This paper proposes a system for private fraud detection on encrypted transactions using homomorphic encryption. Two models, XGBoost and a feedforward classifier neural network, are trained as fraud detectors on plaintext data. They are then converted to models which use homomorphic encryption for private inference. Latency, storage, and detection results are discussed, along with use cases and feasibility of deployment. The XGBoost model has better performance, with an encrypted inference as low as 6ms, compared to 296ms for the neural network. However, the neural network implementation may still be preferred, as it is simpler to deploy securely. A codebase for the system is also provided, for simulation and further development.



\newpage
\onehalfspacing\raggedright 

\section*{\Huge{Acknowledgements}}
Firstly, I would like to thank my supervisor Dr. Ciarán Mc Goldrick for his guidance and valuable feedback during the course of my research. Also, thank you to the Evervault team for introducing me to the concept of homomorphic encryption. I would also like to express my gratitude to my parents, and to my lab partner Katherine, for their continuous support and encouragement.

\tableofcontents
\listoffigures
\listoftables
\newpage






\mainmatter
\chapter{Introduction}
\section{Motivation}

Homomorphic Encryption (HE) is a technology which allows operations to be performed on encrypted data without needing to decrypt it. This concept has long been referred to as the "holy grail" of cryptography \cite{micciancio2010first}, but has historically struggled with speed and accessibility issues. Recent improvements in these areas has led to HE being proposed as a practical solution to real-world problems, such as in the healthcare \cite{prakaashini2021comprehensive} or finance \cite{han2019logistic} domains. I initially read about HE while working at encryption services startup Evervault, and decided to search for some existing problems which it could potentially be a solution for. The intention was to become familiar with the state-of-the-art practices in HE, while solving an important problem. Given the sensitive nature of credit card (CC) data, it stood out as a domain which could benefit from private computation.

CC fraud is the use of someone's payment card fraudulently, typically involving the theft of funds via transfer, or the purchasing of goods. The European Central Bank found that in 2019 alone, the value of fraudulent transactions amounted to over €1.8 billion \cite{ecb2021}. CC fraud is an important issue for banks, as they must make large amounts of payouts to customers each year. It is also a great frustration for customers when one's card has had funds stolen from it, or is frozen due to fraudulent activities. This makes CC fraud detection a vital issue for banks and customers, and a difficult problem to be solved in the financial world. Existing fraud detectors have found success by using rule-based methods or Machine Learning (ML) methods on data surrounding the transactions. This data may include information such as the location at which the transaction was made, or the type of goods or services being purchased.

Although the detectors have resulted in a relative reduction in CC fraud, it has led to the need for users to share additional data, in plaintext form. The upward trend of data breaches motivates a reduction in the amount of sensitive data gathered from users. Additionally, some users with high demands for privacy may not want to share this data with their bank, let alone with potential hackers. This calls for a solution in which user data can be kept secure, while also allowing it to be used in fraud detectors. This paper explores using HE to build a CC fraud detector which meets this need for privacy. As the main focus is on the privacy preservation, the aim is not to make reach state-of-the-art performance at the fraud detection task, but to make a model which performs well as a fraud detector in the HE domain.

ML models in various domains process vast amounts of sensitive user data each day. This provides an additional motivation to produce research which converts standard ML models to models which use HE, as parts of this research can then be used in domains unrelated to CC fraud. Hence, the methods implemented in this paper will be ML based.

An additional motivation is to produce an accessible, open source codebase for this project. It is currently quite difficult to find tutorials for ML inference using HE, or information on how to adapt on previous work. Novel technologies such as HE are accelerated by community contributions. It is hoped that the codebase released with this paper will be valuable for further research.

\section{Approach}

The approach of the paper is to create HE implementations of CC fraud detectors and make latency and storage comparisons with their original plaintext versions. The produced HE models will also be compared, to see which would be most practical to use in the real-world. Comparisons made include transaction encryption latency, model size, transaction size, and inference latency.

The choice of which methods to implement in HE is based on how effective their plaintext versions are at detecting fraud. The number of models chosen for comparison, two in this case, is limited by the amount of time allocated to the project.

\section{Objectives}
\label{sec: objectives}

The objective of this research is to produce a model which, on the selected datasets, achieves the following standards:

\begin{enumerate}
  \item Transaction data is encrypted at 128 bits of security, and operated on using Homomorphic Encryption.
  \item The model achieves high performance detecting the fraudulant transactions in the dataset.
  \item The HE implementation has the same detection performance as the plaintext model.
  \item The inference time of the model is reasonable for use in a service requiring real-time performance.
  \item The time taken to encrypt the model is reasonable for use in a service which may have many users.
  \item The size of the encrypted model is reasonable for use in a service which may have many users.
\end{enumerate}

If possible, multiple models which achieve this standard will be produced. Having models with different characteristics, pros, and cons can open up opportunities for different use cases.

\section{Report Structure}

Section \ref{sec: background} outlines the necessary background information on the topics used, and previous relevant work. Section \ref{sec: pp-cc-fd} details the design of the proposed privacy-preserving CC fraud detection system, with section \ref{sec: implementation} describing the technical details of the implementation. In section \ref{sec: evaluation} the results of CC fraud detection performance are shown, along with latency and storage size benchmarks. Section \ref{sec: further work} discusses the performance of the models, limitations of the system, and proposals of further work on the system. The conclusion of the paper is stated in section \ref{sec: conclusion}.
\chapter{Background}
\label{sec: background}

\section{Credit Card Fraud Detection}

Credit card (CC) fraud is an umbrella term used for any time a payment card (debit or credit) is used against the will of the owner. CC fraud consists of card-present fraud, which is the use of a physical card, and card-not-present fraud, which involves the use of stolen card details for online transactions \cite{ecb2021}. CC fraud detection is the problem of predicting whether a given transaction is fraudulent or legitimate. Designing an effective detection system has been a consistent subject of research for the past three decades \cite{leonard1993detecting,maes2002credit,patidar2011credit,yang2019ffd}.

\section{Transaction Data}

Access to previous transaction data allows for detection models to be based on patterns found in existing data, and then tested on that data. Datasets typically contain a list of transactions in chronological order. Each transaction contains a number of transaction features, such as transaction amount, or product type purchased. They are also labelled as fraudulent or legitimate transactions, represented by 1 or 0 respectively.

Datasets which are made publicly available must undergo a process of anonymization to ensure that information about who made the transactions cannot be inferred. A typical method for anonymization is Principal Component Analysis (PCA) \cite{abdi2010principal}. Some prior works use anonymized, publicly available datasets \cite{ge2020credit}, and others use private commercial datasets \cite{fu2016credit}. A simulator for CC fraud data has also recently been implemented \cite{leborgne2022fraud}, which holds potential to enable further testing of models on generated data.

A small percentage of CC transactions are fraudulent, for instance, only 0.024\% of SEPA transactions \cite{ecb2021}. This skew is also typically present in datasets used for fraud detection, in which often less than 5\% of transactions are fraudulent. Managing this unbalanced data is a challenging problem which must be dealt with by practitioners \cite{dal2014learned}.

\section{Machine Learning Techniques}

In this section, Machine Learning (ML) methods and techniques used in previous works on CC fraud detection are discussed, with two families of ML models being the focus: Decision Trees and Neural Networks. Previous work on CC fraud detetors have also used classical ML algorithms such as Support Vector Machine (SVM) \cite{sahin2010detecting} and logistic regression \cite{sahin2011detecting}. Some ML terminology is required, with definitions provided for Loss Function \ref{def: loss-func}, Overfitting \ref{def: overfitting}, and Feature Engineering \ref{def: feature-engineering}.

\theoremstyle{definition}
\newtheorem{definition}{Definition}

\begin{definition}[Loss Function]
\label{def: loss-func}
When training ML models, the goal of the training process is to decrease the value of a function which usually represents the difference between the current outputs of the model and the intended outputs. This function is known as the loss function.
\end{definition}

\begin{definition}[Overfitting]
\label{def: overfitting}
If an ML model is trained on existing data, this data is known as the \emph{training set}. Overfitting is a term used for when a model learns to perform correctly on the training set, but performs significantly worse when introduced to new data.
\end{definition}

\begin{definition}[Feature Engineering]
\label{def: feature-engineering}
Feature engineering is the process of adding extra features to a dataset with the intention of improving model training.
\end{definition}

\subsection{Decision Trees}

A Classification and Regression Tree \cite{loh2008classification} is a tree structure in which a comparison is made at each \emph{node}, and the \emph{leaf node} reached contains a result. These are often referred to as CARTs, regression trees or decision trees. For simplicity, in this paper are referred to as \emph{trees}. In 2007, Shen et al. attempted to use a single tree to detect CC fraud \cite{shen2007application}, but it was outperformed by other models.

\begin{figure}[H]
      \centering
      \includegraphics[width=\textwidth]{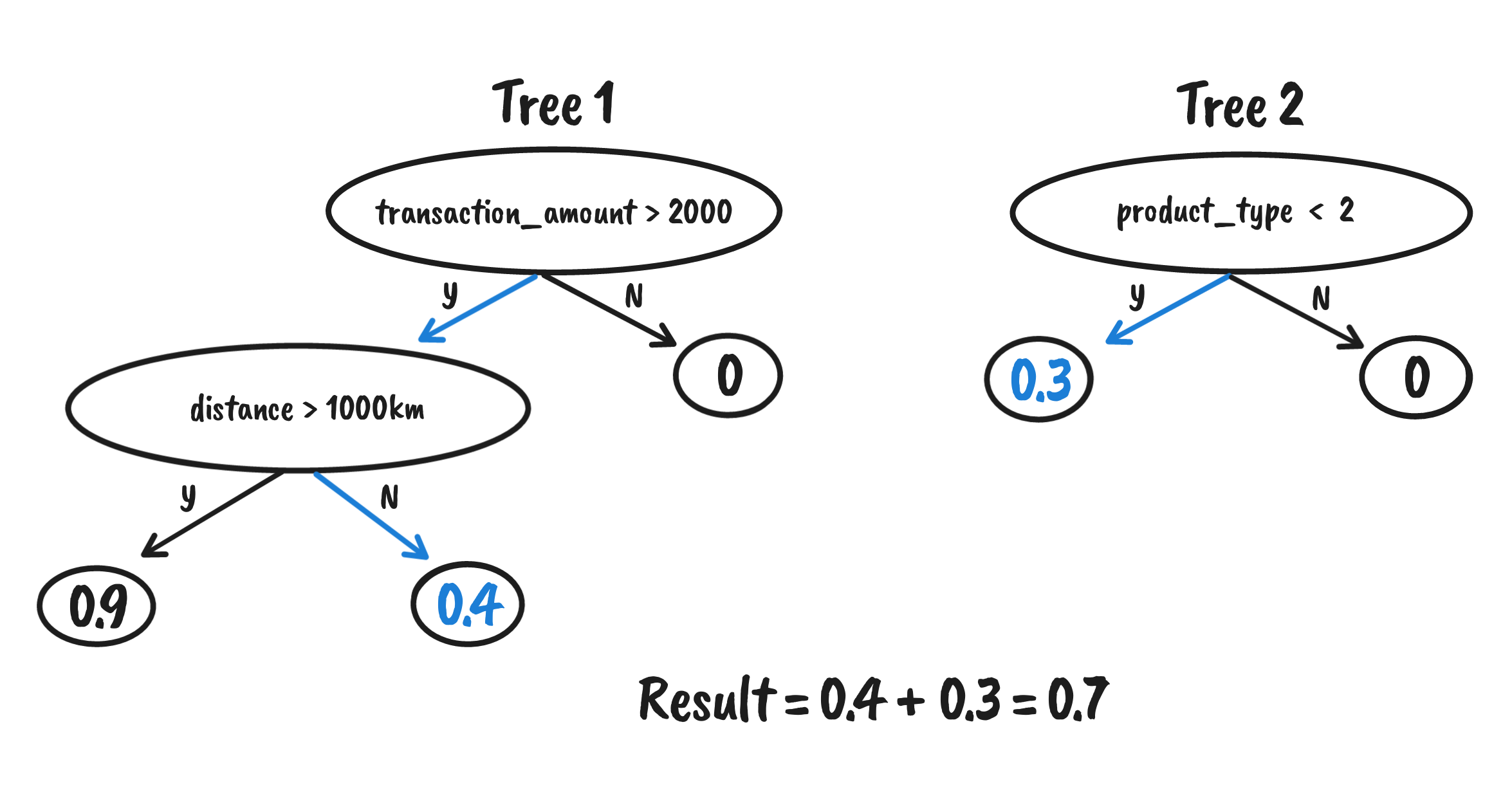}
      \caption{Decision tree ensemble for predicting probability of CC fraud}
      \label{fig: tree-ensemble}
\end{figure}

An ensemble of trees is when the result of multiple trees is combined to produce the output. Random Forest, an ensemble method, was proposed as a method for CC fraud detection, but was unable to adequately handle the unbalanced data \cite{xuan2018random}. Gradient boosting \cite{friedman2000additive} is a method which has been used to improve the training of the trees. Models using gradient boosting have found success for CC fraud detection, such as XGBoost \cite{priscilla2020influence} and LightGBM \cite{ge2020credit}. An example of how a tree ensemble model could be used to predict a probability for if a transaction is fraudulent is shown in Figure \ref{fig: tree-ensemble}.

\subsection{Neural Networks}

An Artificial Neural Network (NN) is an ML model which processes data using a combination of functions called \emph{neurons}, which calculate a learned linear function on its input. The simplest form of NN is called a feedforward NN. In a feedforward NN, numerical inputs are passed through a series of \emph{layers} to produce an output. The \emph{depth} of a feedforward NN corresponds to the number of layers it has. Each layer contains a set of neurons. The output of each layer passes through an \emph{activation function} before being passed to the next layer. Activation functions are used to introduce non-linearity to the model. The most popular activation function is the Rectified Linear Unit (ReLU) \cite{xu2015empirical}.

\begin{figure}[H]
      \centering
      \includegraphics[width=0.6\textwidth]{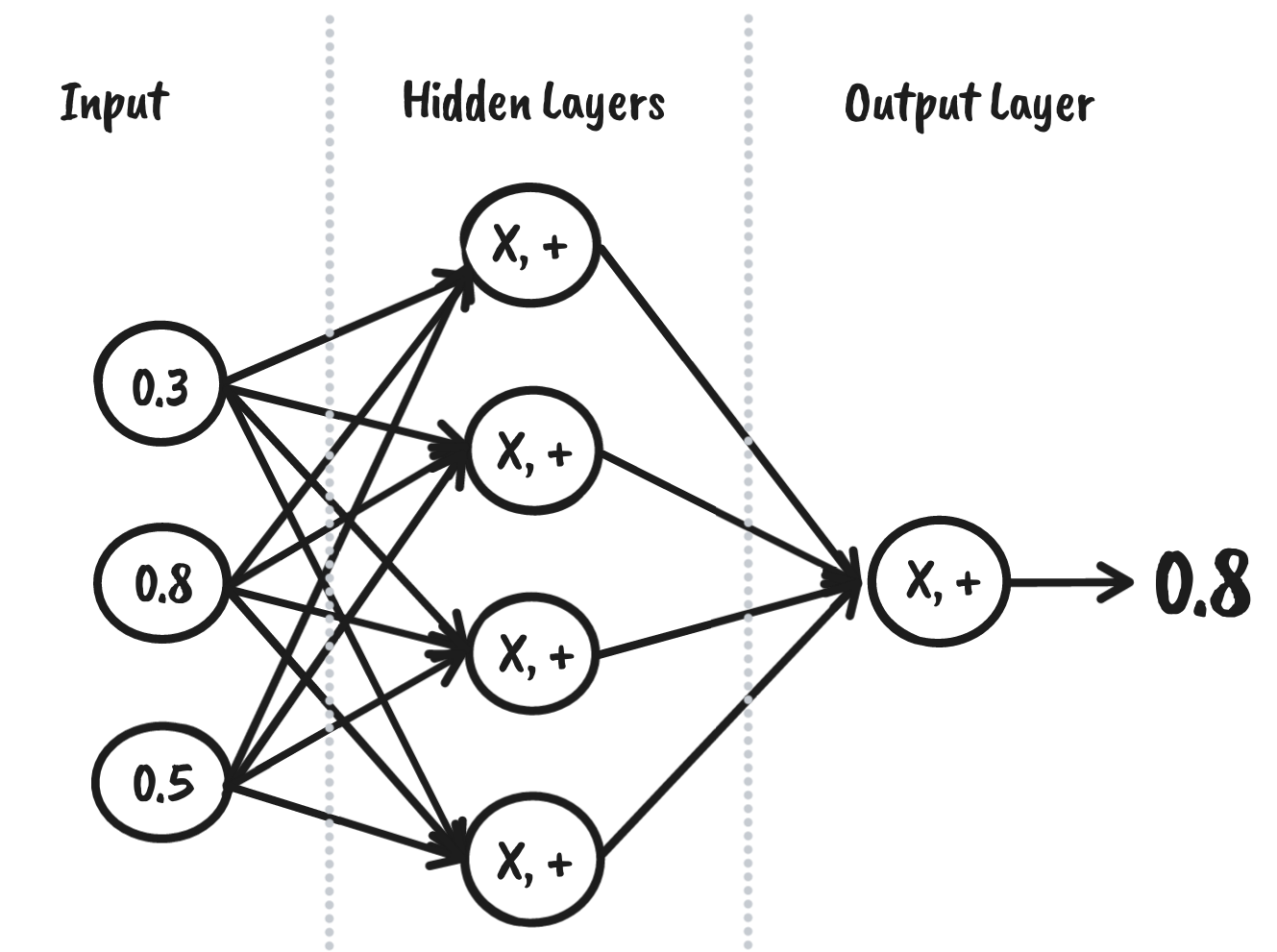}
      \caption{Feedforward Classifier NN for predicting probability of CC fraud}
      \label{fig: feedforward}
\end{figure}

A Classifier NN is one which produces a single output which is either a binary value or a continuous probability. This architecture was used in the first implementation of an NN for CC fraud detection \cite{ghosh1994credit}. An example of how a feedforward classifier NN could be used to predict a probability for if a transaction is fraudulent is shown in Figure \ref{fig: feedforward}.

Other types of NNs previously used for CC fraud detection include Autoencoder \cite{aleskerov1997cardwatch,al2019credit}, Convolutional Neural Network (CNN) \cite{fu2016credit} and Long Short Term Memory \cite{jurgovsky2018sequence}.

\section{Homomorphic Encryption}

\begin{figure}[H]
      \centering
      \includegraphics[width=0.8\textwidth]{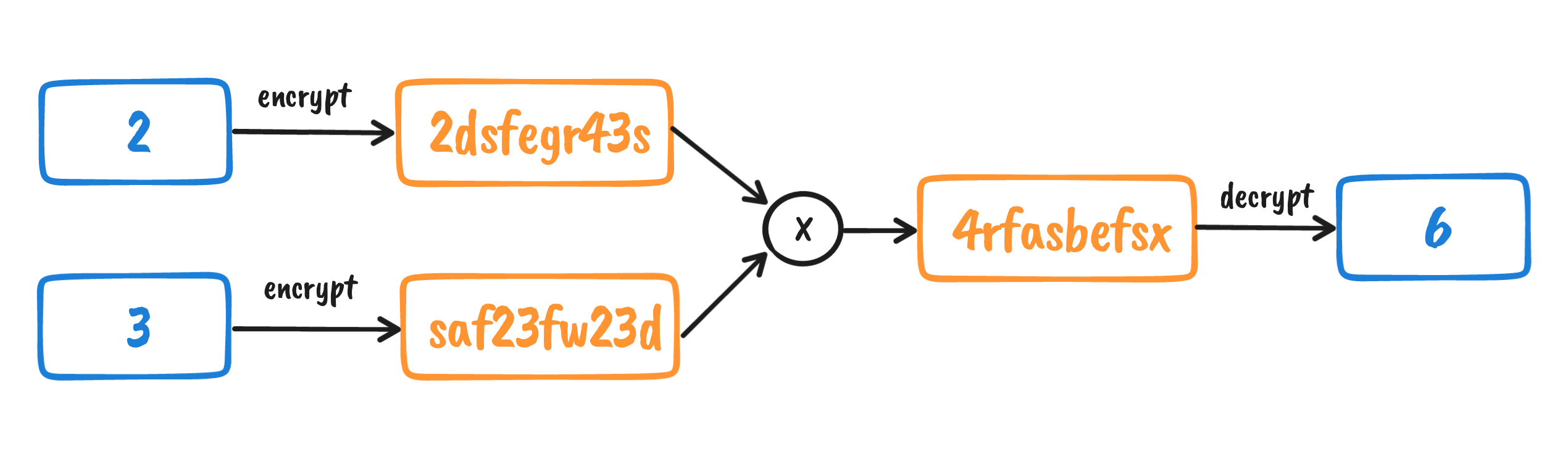}
      \caption{Secure multiplication using Homomorphic Encryption}
      \label{fig: homomorphic-multiplication}
\end{figure}

Encryption is the process of converting data into another form such that the original value can only be revealed by the holder of a \emph{secret key}. The encrypted form is called a \emph{ciphertext} and the plain form is called a \emph{plaintext}. Homomorphic Encryption (HE) is a way of encrypting data such that some or all standard operations (addition, multiplication, division etc.) can be performed using ciphertexts, with the decrypted result showing the correct value. An example of multiplication using HE is shown in Figure \ref{fig: homomorphic-multiplication}, as a demonstration of this concept.

HE was originally proposed by Rivest et. al in 1978 \cite{rivest1978data}, with Fully Homomorphic Encryption (FHE) being the end goal. FHE schemes are those which allow for any operation to be evaluated any number of times. The FHE problem was cracked in 2009 by Gentry \cite{gentry2009fully}, but choosing an FHE scheme is not always optimal for performance. \emph{Partially} or \emph{leveled} HE schemes are often chosen, which have limitations on, respectively, the type of operations allowed or the number of operations allowed, but may provide lower latency. A comprehensive survey by Acar et al. contains information on the history and specifics of various schemes \cite{acar2018survey} and also found that the dominant topics of research for HE are in the health and finance domains.

Definitions for Symmetric Encryption \ref{def: symmetric} and Asymmetric Encryption \ref{def: asymmetric} are given to aid in understanding the uses of the HE schemes discussed.

\begin{definition}[Symmetric Encryption]
\label{def: symmetric}
In a symmetric encryption scheme a single secret key is used to encrypt and decrypt.
\end{definition}

\begin{definition}[Asymmetric Encryption]
\label{def: asymmetric}
In an asymmetric encryption scheme a public key and a secret key are generated. The public key can be safely sent to anyone, and anyone can use it to encrypt values. However, only the holder of the secret key can decrypt.
\end{definition}

\section{Decision Tree Inference using Homomorphic Encryption}

The first implementation of a Decision Tree using HE was by Bost et al. in 2014 \cite{bost2014machine}. They used the Quadratic Residuosity scheme \cite{goldwasser2019probabilistic} and a leveled HE scheme implemented by HElib \cite{halevi2014algorithms}, with optional ensembling using the Paillier scheme \cite{paillier1999public}. Inference time of approximately 1 second, depending on security parameters and tree depth.

In 2015, Khedr et al. demonstrated an encrypted tree ensemble model \cite{khedr2015shield} which used a Ring Learning With Errors variant of the GSW scheme \cite{gentry2013homomorphic}. They achieved a much better inference time of 0.027s.

In 2020, X. Meng \& J. Feigenbaum released Privacy-Preserving XGBoost Inference (PPXGB) \cite{meng2020privacy}, which uses the Order Preserving Encryption (OPE) \cite{boldyreva2009order} scheme and Paillier in an encrypted implementation of XGBoost. They demonstrated an inference time of 0.32s.

\section{Neural Network Inference using Homomorphic Encryption}

Significantly more research has been done on HE implementations of NNs than Decision Trees. This is likely due to the trend of deep learning, and the applicability of NNs to problems in various domains. CryptoNets \cite{gilad2016cryptonets} in 2016 was the first implementation, which used the YASHE scheme \cite{bos2013improved} and a square activation function. CryptoNets performed well on its classification task, but still had a high latency of 250 seconds.

Latency improvements have since been made by using a GPU in HCNN \cite{al2018towards}, and by more efficient encoding in LoLa \cite{brutzkus2019low}, which both used the BFV scheme \cite{fan2012somewhat}. nGraph-HE2 \cite{boemer2019ngraph} then achieved state-of-the-art throughput using the CKKS scheme \cite{cheon2017homomorphic}, which is a fast HE scheme for approximate computation.

Clet et al. made a comparison on the BFV, CKKS and TFHE \cite{chillotti2020tfhe} schemes and found that the CKKS scheme was the fastest for inference using a small architecture \cite{clet2021bfv}. Some works have used polynomial approximations of activation functions such as ReLU \cite{chabanne2017privacy} for improved performance in exchange for added latency.

\section{Alternatives to Homomorphic Encryption}

\subsection{Different methods for private computation}

Alternative methods to HE for private computation have also been proposed. A Trusted Execution Environment (TEE) \cite{sabt2015trusted} is a compute environment in which code can be run in a confidential manner and the code running can be remotely attested. Secure Multi-Party Computation (MPC) \cite{goldreich1998secure} is a method for sharing the compute of a function between parties, without any single party knowing the full information.

\subsection{Anonymization}

If data can be anonymized in such a way that detection models using the anonymized data reach the same level of performance of models which use private data, then simply anonymizing the data would result in privacy-preserving detection.

\subsection{Using Non-Sensitive Data}

Similarly to anonymization, if adequate performance can be reached using data which is deemed non-sensitive, then there is no need for privacy-preserving methods.

\section{Related Work}

HE for CC fraud detection was explored in two previous works. Canillas et al. used the original HE encrypted tree implementation \cite{canillas2018exploratory} to implement an encrypted single-tree CC fraud detector, with a latency of approximately 1 second.  Vazquez-Saavedra et al. used the CKKS scheme to implement an encrypted SVM model. This achieved very low latency (<1ms in some cases), but the encrypted model did not have the same detection performance as the plaintext \cite{vazquez202154}.
\chapter{Privacy-Preserving Credit Card Fraud Detection}
\label{sec: pp-cc-fd}

This section proposes a system for private CC fraud detection using HE, including use cases, the methods implemented, and the system architecture of each method.

\section{Use Cases}
\label{sec: introduction-use-cases}

It is important to understand the use cases of the proposed system to ensure that it has practical applications. Each use case has (1) Host, which is the party hosting the HE CC fraud detetor, and (2) Client, which is the party sending an encrypted transaction to Host in order to find out if the transaction(s) is fraudulent or not.

\subsection{Customer \& Bank}
\label{sec: use-cases-customer-bank}

In this use case, a bank is Host. Customers of the bank are Clients. A customer attempts to make a CC transaction. Before the transaction goes through, its features are encrypted and sent to Host for private fraud detection. If the transaction is predicted as legitimate, the transaction goes through. If not, the transaction is blocked.

Every customer of the bank could potentially be a Client, or perhaps only a subset of customers who are willing to accept extra transaction latency or payment costs, in exchange for a higher guarantee of privacy. A limitation on this use case is discussed in Section \ref{sec: limitation-encrypted}.

\subsection{Bank \& Third-Party}

In this use case, a bank is Client. Host is any third-party which has created a (plaintext) CC fraud detection model which has better detection performance than the model currently available to the bank. The bank does not trust the third-party with plaintext customer data, so the third-party hosts a HE implementation of their model. The bank can then send encrypted transaction data to the third-party to make use of the superior fraud detector. The third-party could be another bank, or a company set up to provide this service. Note that in this use case, the Client bank could theoretically store their customers transaction data in plaintext, which would not protect from data breaches.

\section{Methods Implemented}

A Decision Tree model (XGBoost) and a NN model (Feedforward Classifier) were both selected for use in the system. Selecting from two different families of ML algorithms can demonstrate the usability of each and can reveal features of each which may be more suited to certain use cases. If both are performant, it also opens up further opportunities for a combination via ensemble.

\subsection{HE XGBoost}

XGBoost was chosen due to the high performance of its plaintext implementation on CC fraud detection tasks. In addition, the open-source implementation of PPXGB\footnote{https://github.com/awslabs/privacy-preserving-xgboost-inference} meant there would be little experimentation required for successfully converting a model to HE.

\begin{figure}[H]
      \centering
      \includegraphics[width=\textwidth]{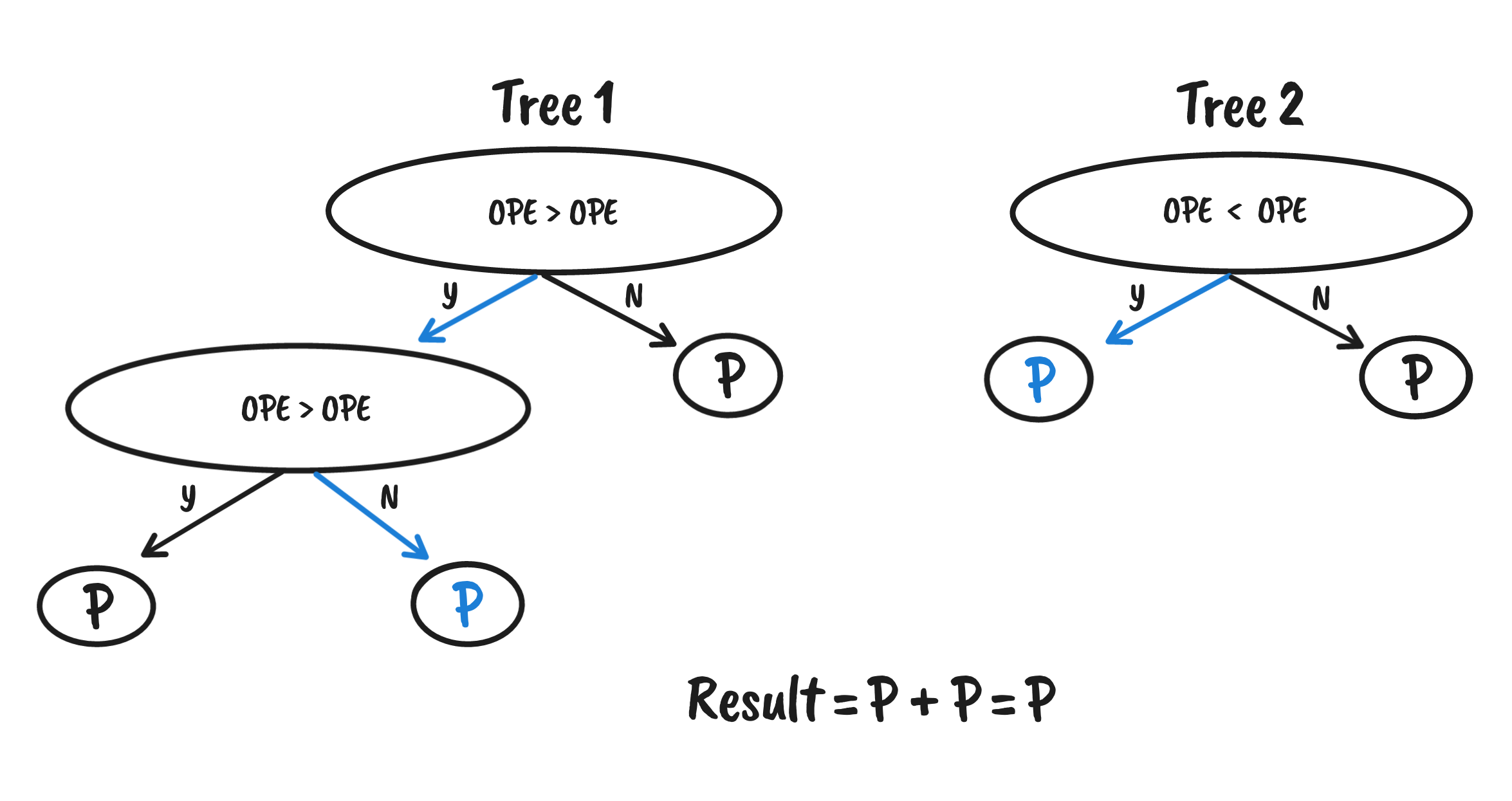}
      \caption{XGBoost model after conversion to HE. OPE represents a value encrypted using an Order Preserving Encryption scheme, P represents a value encrypted using the Paillier encryption scheme.}
      \label{fig: he-xgboost}
\end{figure}

An example of what the XGBoost model previously shown in Figure \ref{fig: tree-ensemble} would look like after conversion to HE using PPXGB is shown in Figure \ref{fig: he-xgboost}.

\subsection{HE Feedforward Classifer}

A Feedforward Classifier NN model was chosen due to its previous success for CC fraud detection, and its simple architecture. The HE implementation used is the CKKS scheme, as it was found to be have the fastest inference time for a similar architecture \cite{clet2021bfv}, and open-source implementations are supported in various programming languages. The CKKS scheme works well for NNs, as it has fast latency by using approximate arithmetic. As Feedforward Classifer NNs produce a fuzzy output, using approximate arithmetic can result in a slightly different output, but show the same prediction after conversion to binary.

\begin{figure}[H]
      \centering
      \includegraphics[width=0.6\textwidth]{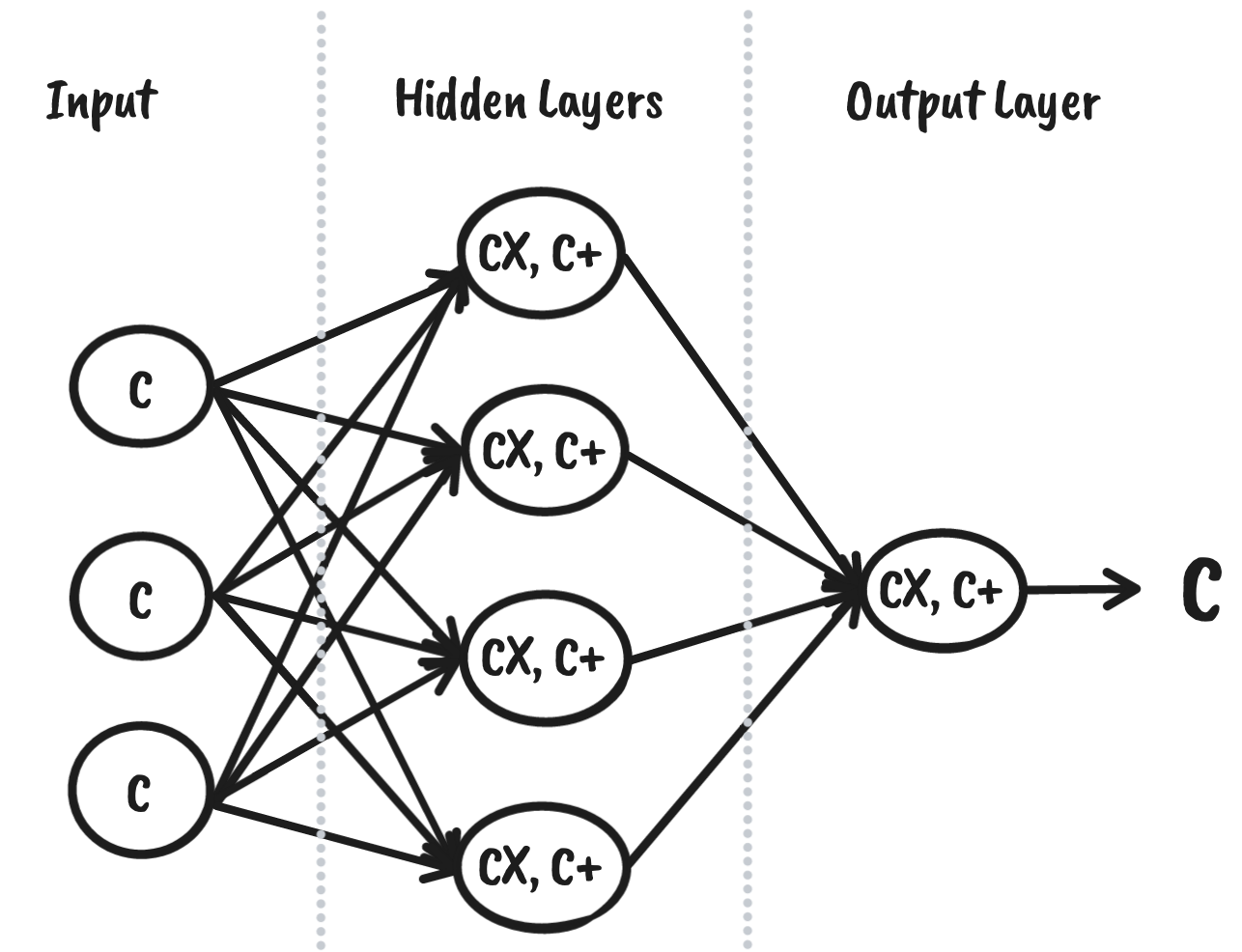}
      \caption{CKKS implementation of feedforward NN model. C represents a value encrypted using the CKKS scheme. CX,C+ represent multiplications and additions using the CKKS scheme.}
      \label{fig: ckks-feedforward}
\end{figure}

An example of a CKKS implementation of the NN previously shown in Figure \ref{fig: feedforward} is shown in Figure \ref{fig: ckks-feedforward}.

\section{System Architecture}
\label{sec: sys-arch}

\begin{figure}[H]
      \centering
      \includegraphics[width=\textwidth]{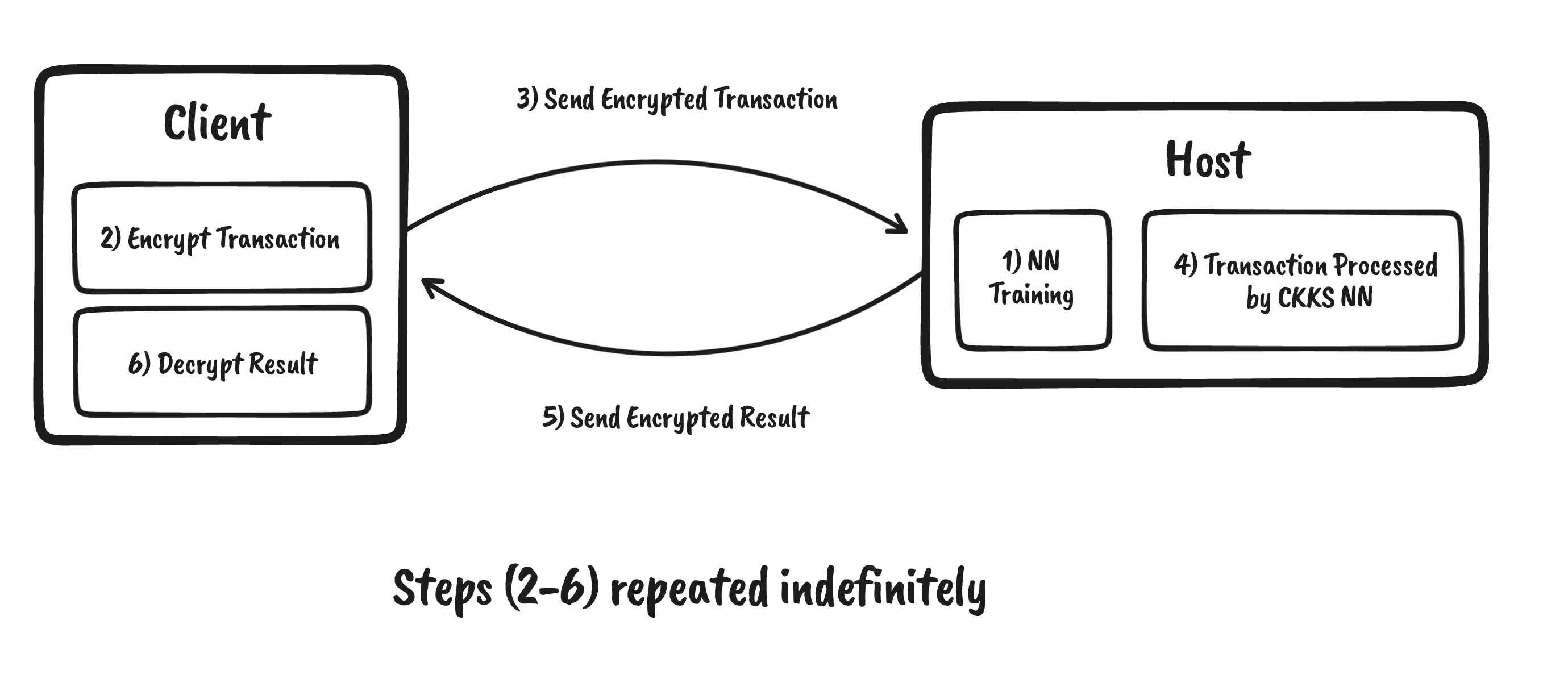}
      \caption{Sequence of operations in CC Fraud detection system using CKKS NN}
      \label{fig: nn-sequence-diagram}
\end{figure}

The sequence of events for a CC fraud detection system using the HE NN is shown Protocol \ref{alg: he-nn}, with a visualisation shown in Figure \ref{fig: nn-sequence-diagram}. Once step 1 is completed, any Client can take part in steps 2-6 an infinite number of times.

\begin{algorithm}[H]
\caption{HE NN for CC fraud detection}
\label{alg: he-nn}
\begin{algorithmic}
\STATE $H$: Host
\STATE $C$: Client
\end{algorithmic}

\algsetup{linenosize=\small,linenodelimiter=.}
\begin{algorithmic}[1]
\STATE $H$ trains NN model on plaintext transaction data.
\STATE $C$ encrypts a transaction.
\STATE $C$ sends encrypted transaction to $H$.
\STATE $H$ processes encrypted transaction using HE NN, producing encrypted result.
\STATE $H$ sends encrypted result to $C$.
\STATE $C$ decrypts to view the result.
\end{algorithmic}
\end{algorithm} 

\begin{figure}[H]
      \centering
      \includegraphics[width=\textwidth]{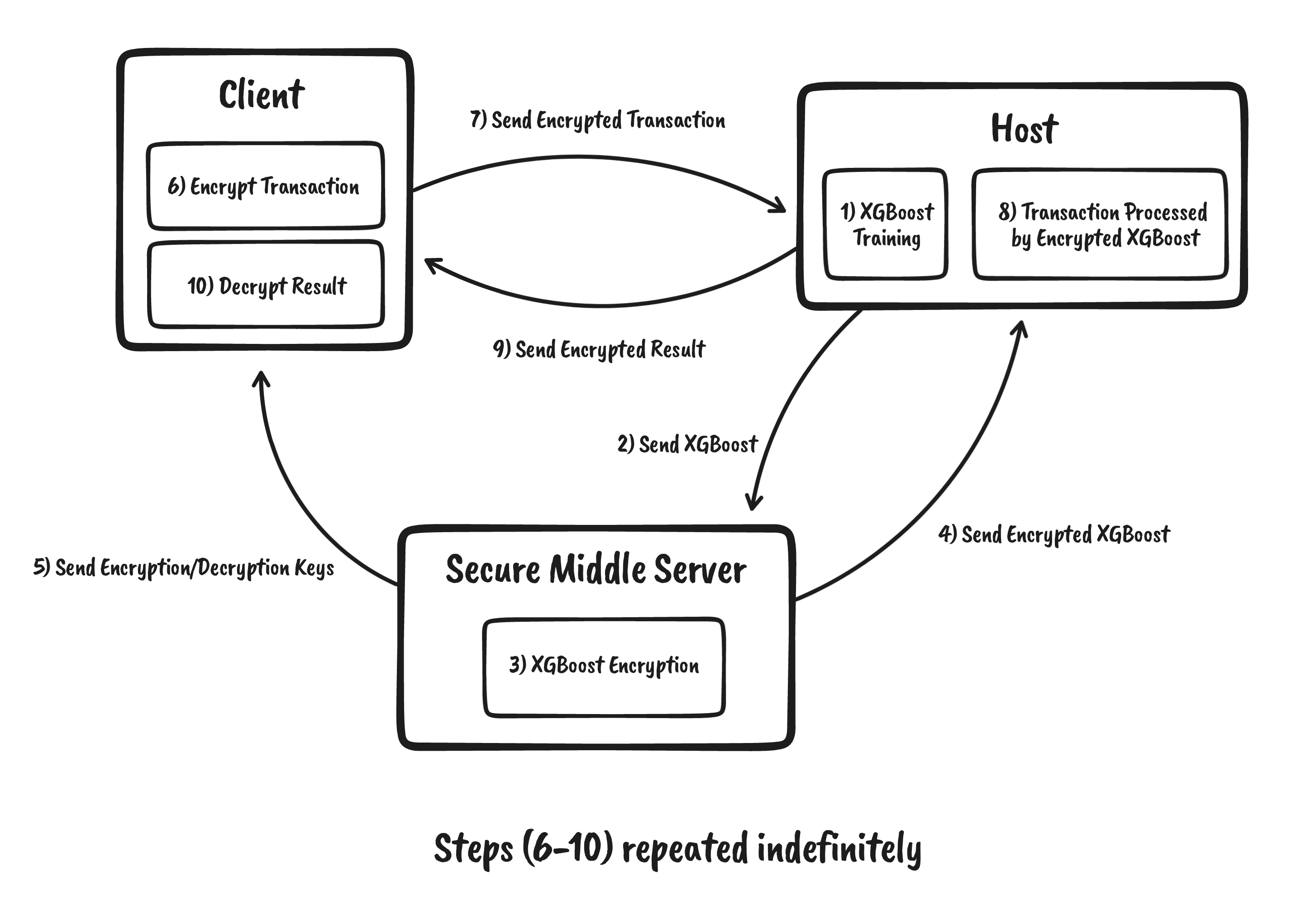}
      \caption{Sequence of operations in CC Fraud detection system using encrypted XGBoost}
      \label{fig: xgboost-sequence-diagram}
\end{figure}

The sequence of events for a CC fraud detection system using the HE XGBoost model is shown in Protocol \ref{alg: he-xgboost}, with a visualisation shown in Figure \ref{fig: xgboost-sequence-diagram}. As PPXGB uses symmetric encryption, the model must be encrypted in a \emph{Secure Middle Server}. In a production setting, this middle server would be run in a TEE. Steps 2-5 must be done once \emph{per Client}. Once completed, they can take part in steps 6-10 an infinite number of times.

\begin{algorithm}[H]
\caption{HE XGBoost for CC fraud detection}
\label{alg: he-xgboost}
\begin{algorithmic}
\STATE $H$: Host
\STATE $C$: Client
\STATE $SMS$: Secure Middle Server
\end{algorithmic}

\algsetup{linenosize=\small,linenodelimiter=.}
\begin{algorithmic}[1]
\STATE $H$ trains XGBoost model on plaintext transaction data.
\STATE $H$ sends XGBoost model to $SMS$.
\STATE $SMS$ generates $K$ and encrypts XGBoost model.
\STATE $SMS$ sends HE XGBoost model to $H$.
\STATE $SMS$ sends keys to $C$.
\STATE $C$ encrypts a transaction using keys.
\STATE $C$ sends encrypted transaction to $H$.
\STATE $H$ processes encrypted transaction using HE XGBoost, producing encrypted result.
\STATE $H$ sends encrypted result to $C$.
\STATE $C$ decrypts to view the result.
\end{algorithmic}
\end{algorithm} 

\section{Datasets Chosen}

The datasets chosen for model training are the Vesta and ULB datasets. The reasons for choosing these is that they are the two largest public datasets available, and have both been widely used in previous work since their respective releases.

No extra feature engineering was done on the datasets, as the focus of this paper is not on analysing the datasets. Using the datasets in their existing form with little pre-processing demonstrates how the models can be implemented directly on existing data.

Both of the datasets used have undergone some level of anonymization. It may seem impractical to use them in a privacy-preserving way, as they don't contain any private data. However, they still work as a proof-of-concept for privacy-preserving inference. The reason for this is that anonymizing the data can only result in a level of entropy which is \emph{less than} or \emph{equal to} the amount of entropy in the original data. Hence, if the level of entropy in the anonymized data can be used to accurately predict if transactions are fraudulent, it follows that there would be enough entropy present in the original data to reach at least the same level of performance.

\subsection{"Credit Card Fraud Detection" (ULB)}

The ULB dataset\footnote{https://www.kaggle.com/datasets/mlg-ulb/creditcardfraud} contains 284,807 transactions, 492 of which are labelled as fraudulent (0.172\%). Each transaction has 28 numerical features V1-V28, along with the relative time since the first transaction and a label of whether it was fraudulent or legitimate. PCA has been applied to the dataset before release for anonymization, with V1-V28 being the first 28 principal components.

\begin{figure}[H]
      \centering
      \includegraphics[width=\textwidth]{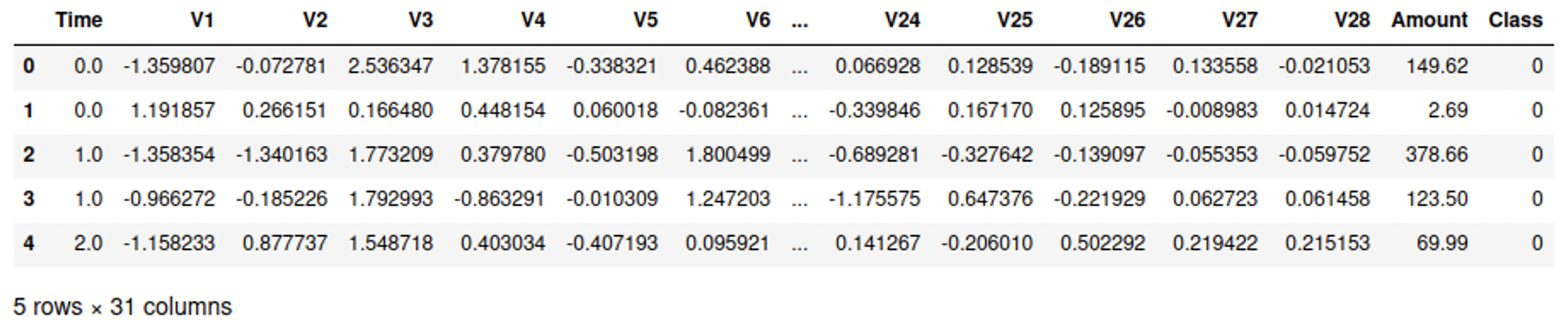}
      \caption{ULB dataset sample}
      \label{fig: ulb-sample}
\end{figure}

\subsection{"IEEE CIS Fraud Detection" (Vesta)}
\label{sec: comp}

\begin{figure}[H]
      \centering
      \includegraphics[width=\textwidth]{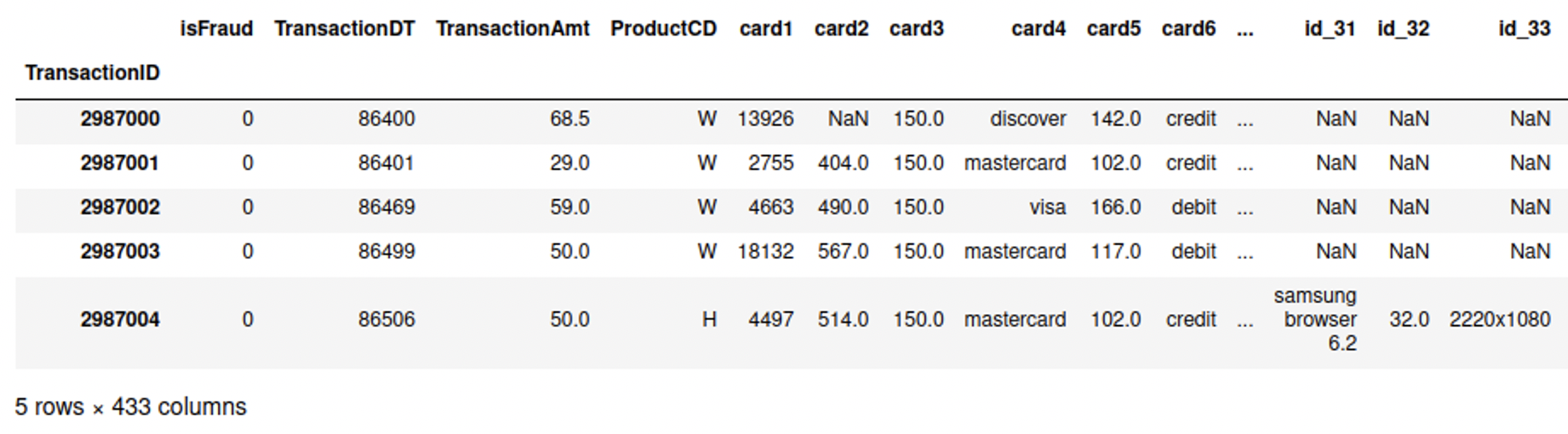}
      \caption{Vesta dataset sample}
      \label{fig: vesta-sample}
\end{figure}

The Vesta dataset was released for use in the IEEE CIS Fraud Detection competition\footnote{https://www.kaggle.com/competitions/ieee-fraud-detection/}. It contains 590,540 transactions, 20,663 of which are fraudulent (3.5\%). Each transaction has 431 features (400 numerical, 31 categorical), along with the relative timestamp and a label of whether it was fraudulent or legitimate. For anonymization purposes, the names of the identity features have been masked, along with the names of the extra features engineered by Vesta. Some features are \emph{sparse}, meaning the feature is empty empty or \emph{Not a Number} (NaN) in many transactions. Figure \ref{fig: vesta-sample} shows the first 5 rows in the dataset.

\chapter{Implementation}
\label{sec: implementation}
This section explains the technical details of the implementation. The code written for this implementation is open-sourced in a GitHub repository\footnote{https://github.com/davidnugent2425/he-cc-fraud-detection.git}. All scripts and notebooks mentioned are located in the \verb|src| directory.

\section{Core Language \& Tools}

\subsection{Python}

Python 3 was chosen as the programming language for the implementation, as this is the language used in PPXGB, and the CKKS scheme also has Python implementations.

\subsection{Weights \& Biases}

Weights \& Biases (W\&B) \cite{wandb} is a tool which aids in training ML models, via effective logging and visualisation of training runs. Including W\&B in the training pipeline allows for logging of metrics to ensure models are training properly and to keep track of the performance of each model.

W\&B also provides a free service for doing \emph{Sweeps} of model \emph{hyperparameters}. These are the parameters associated with the structure of a model such as the size of a model layer, or the number of decision trees in a model. A Sweep is a sequence of training runs, each using different hyperparameters. The results of each training run are logged to W\&B for comparison. The choice of which hyperparameters used in each training run can be either a random choice, or using a bayesian method to try converge toward better performance on a chosen metric.

\subsection{Jupyter Notebook}

Jupyter Notebooks \cite{granger2021jupyter} are interactive Python notebooks which allow researchers to display their results in an interactive manner, by running inline code alongside standard markdown text. They are used in the simulation to aid in understanding which code would be run by each of the parties involved.

\subsection{Model Training Packages}

The XGBoost Python Package \cite{chen2016xgboost} is used for training the XGBoost models. PyTorch \cite{NEURIPS2019_9015} was found to be suitable for converting models to HE, as it uses an object oriented approach.

\subsection{TenSEAL}

TenSEAL \cite{benaissa2021tenseal} is a library which contains Python bindings for the C++ CKKS scheme implementation in Microsoft SEAL \cite{sealcrypto}. TenSEAL have also provided tutorials\footnote{https://github.com/OpenMined/TenSEAL/tree/main/tutorials} on converting plaintext NNs to HE models, which was helpful for writing this implementation.

\section{Model Training}

Each time a dataset is used for training, the first 65\% of transactions are used as the \emph{training set}, the next 15\% are used as the \emph{validation set}, and the final 20\% are used as the test set. The model is first trained on the training set with various hyperparameters. The choice of hyperparameters is then based on performance on the validation set. Final results are then taken on the test set. 

The purpose of splitting the dataset in this manner is to ensure the resulting model performs well across different distributions, i.e. to prevent overfitting. If the model performs well on the validation and test sets, we can be confident that it would perform well on further new transactions.

Two scripts \verb|train_xgb.py| and \verb|train_nn.py| are provided to allow for simple training of each model. The \verb|dataset| flag can be used when running the scripts to choose the dataset to train on. The \verb|wandb| flag can be used to log the metrics from the training run to W\&B. Logging the training run is important to ensure that the loss function is converging towards a minimum on the training set. If it is not converging, there is likely a problem with either the loss function, or perhaps the learning rate is too high. The full output of an XGBoost training run on the ULB Dataset is shown in Appendix \ref{fig: sim-ulb-training}.

\subsection{Undersampling}

As the datasets are highly skewed towards legitimate transactions, improvements may be found in a model's performance by using a method called \emph{undersampling} \cite{dal2015calibrating}. In undersampling, a subset of the training set is selected which increases the total percentage of fraudulent transactions. The intention of undersampling is to achieve better \emph{recall} on the fraudulent transactions, as the model is less skewed toward labelling transactions as legitimate. A hyperparameter \verb|undersampling_num_negatives| is implemented in the training pipeline which determines the number of legitimate transactions included in the training set.

\subsection{Use of Categorical Variables}

NNs typically use learned embedding layers to handle categorical features \cite{guo2016entity}. However, little work has been done on HE implementations of embedding layers. XGBoost does not currently support any extra optimizations for the use of categorical variables. For these reasons, the categorical features in the Vesta dataset are converted to numerical data.

For conversion to numerical, the pre-processing steps given in the starter code\footnote{https://www.kaggle.com/code/inversion/ieee-simple-xgboost/notebook} provided by the organisors of the competition mentioned in Section \ref{sec: comp}. A simple label encoding converts the categorical features to a corresponding number and the NaN features are replaced with -999.

\subsection{Choosing Model Hyperparameters}

The \verb|wandb-sweep| flag can be used to perform a hyperparameter Sweep of the model. These Sweeps were performed on each model to determine the hyperparameters which performed best on the validation set.

Table \ref{tab: xgboost_hyperparams} shows the ranges of hyperparameters used in the Sweeps for XGBoost on each dataset. \verb|max_depth| controls the maximum depth of each decision tree, and \verb|num_estimators| controls the number of decision trees in the model. Note that \verb|max_depth| was hardcoded as 7 for the Vesta dataset, as it was found that using trees any deeper resulted in impractical model encryption times.

\begin{table}[H]\centering
\caption{XGBoost Hyperparameter Selection}\label{tab: xgboost_hyperparams}
\scriptsize
\begin{tabular}{lrrrr}\toprule
\textbf{Dataset} &\textbf{max\_depth} &\textbf{num\_estimators} &\textbf{undersampling\_num\_negatives} \\\midrule
ULB &6-12 &50-150 &400-2600 \\
Vesta &7 &50-150 &400-300,000 \\
\bottomrule
\end{tabular}
\end{table}

Table \ref{tab: nn_hyperparams} similarly shows the hyperparameters considered for the NN model. \verb|hidden_layer_size| controls the number of neurons in the hidden layer of the model, while \verb|pos_weight| controls an option in the training loss function (BCEWithLogitsLoss), which gives successfully labelled fraudlent transactions an extra weighting. A square activation function was used, as standard activations such as ReLU do not have direct implementations in CKKS, and it only adds a multiplicative depth of 1 per layer.

\begin{table}[H]\centering
\caption{Neural Network Hyperparameter Selection}\label{tab: nn_hyperparams}
\scriptsize
\begin{tabular}{lrrrr}\toprule
\textbf{Dataset} &\textbf{hidden\_layer\_size} &\textbf{pos\_weight} &\textbf{undersampling\_num\_negatives} \\\midrule
ULB &1-5 &400-2600 \\
Vesta &1-5 &400-300,000 \\
\bottomrule
\end{tabular}
\end{table}

\section{Conversion to Homomorphic Encryption}

\subsection{HE XGBoost}
\label{sec: he-conversion-xgboost}

PPXGB encrypts the model by iterating through each of the trees and encrypting each node in the tree recursively using OPE. Note: An additional requirement when using the PPXGB model is that the max and min values in the dataset must be known in order to use the OPE scheme. Once a leaf node is reached, the encryption scheme switches to Paillier. A HMAC is used as a pseudorandom function to obfuscate the names of the features in a transaction \cite{gavzi2014exact} for additional privacy.

An improvement was made to PPXGB to allow for parallelising the tree encryption. As each tree can be encrypted independently, the encryption can be spread across multpile CPU cores. In this implementation, the number of cores can be specified using the \verb|cores| flag. The list of trees is then evenly split into \verb|cores| lists, and encryption for each list is run on a seperate core. In theory, if the number of trees $T$ is greater than the number of cores chosen $C$, there should be a speedup by a factor of $C$ over the original PPXGB function. In practice, thread management adds a cost which reduces the speedup.

\subsection{HE Neural Network}

A layer in a NN performs a matrix mutiplication of the layer input with the layer \emph{weights}, and then adds the \emph{bias} of the layer to produce the output. This evaluate the output of all the neurons in a layer in one operation. The CKKS scheme supports ciphertext-plaintext multiplication. This means the NN layer weights and biases do not need to be encrypted to be used as HE layer weights. Converting from a plaintext NN to a HE NN involves using a CKKS matrix multiplication on the encrypted input with the plaintext layer weights, followed by adding the plaintext layer bias.

A comparison between the model classes showing the corresponding layer operations is shown in Appendix \ref{fig: he-layers-comp}. A debug option was also implemented to display the output of each layer, as shown in Appendix \ref{fig: nn-he-nn-comp}.

\section{HE Inference}

For the HE XGBoost model, Client must be supplied with the OPE key, the HMAC function, and the Paillier private key. Client hashes the transaction feature names with the HMAC, then encrypts the transaction features using the OPE key. This ciphertext is then sent to Host. When the encrypted result is received from Host, Client uses the Paillier private key to decrypt and view the result.

For the HE NN model, Client generates a CKKS public/private key pair. The transaction features are encoded as a CKKS vector and encrypted using the public key. This ciphertext is then sent to Host, along with the encryption context used, so Host can use the same context for the CKKS layer operations. The context could either be the same for every Client, or perhaps some Clients could choose higher levels of security. When the encrypted result is received from Host, Client uses the private key to decrypt and view the result.

\section{Security}

PPXGB uses the python-paillier library \cite{PythonPaillier} for its implementation of the Paillier scheme, which chose a default key length of 3078 to provide 128 bits of security. The symmetric OPE key and the seed used for the HMAC are both 128 bits in length, providing 128 bits of security.

Microsoft SEAL requires that the security context is set up with 128 bits of security by default, by ensuring their encryption contexts are in accordance with the Homomorphic Encryption Security Standard \cite{HomomorphicEncryptionSecurityStandard}. This ensures that if the HE NN runs, it will have 128 bits of security.

\section{Simulation}

A simulation is provided to demonstrate the order of events in a Client-Host use case as described in Section \ref{sec: introduction-use-cases} using the models chosen for this implementation. Additionally, a Secure Middle Server is used for encrypting the XGBoost models. The \verb|Client.ipynb|, \verb|Host.ipynb|, and \verb|Secure Middle Server.ipynb| notebooks represent the respective entities.

Host begins by training XGBoost on the UML and Vesta datasets. Then the NN model is trained on the UML dataset. In each case, the hyperparameters used are those chosen in Tables \ref{tab: xgboost_hyperparams} \& \ref{tab: nn_hyperparams}. The plaintext models are saved to the Host Files.

Next, the Secure Middle Server loads the XGBoost models from Host Files. Encryption/decryption keys are generated, and each model is encrypted. The encrypted models are saved to Host Files, and the keys are saved to Client Files. In practice, the Secure Middle Server would not have direct access to these file systems, and the transmission of the models and keys would be done over secure channels. For simplicity of the simulation, access to both file systems is granted.

Host provides Client with the specification for encoding and encrypting their transaction data, so that it is sent to Host in the correct format for input to the HE Model. Now that Host and Client have the necessary models and keys, the standard protocols for transaction inference described in Section \ref{sec: sys-arch} is done over HTTP.

Each part of the inference process is timed for comparison: encryption time, inference time, and the round trip time from the encryption of the input to decryption of output. This round-trip-time is the most accurate representation of how fast the system performs. However, as this simulation is designed to be run on the same machine, the typical latency involved with transmission over the internet is not included. Appendix \ref{fig: sim-client-host} shows an example of the Client and Host interaction in the Simulation.
\chapter{Results}
\label{sec: evaluation}
In this section the results of HE conversion testing, fraud detection, latencies and storage sizes are presented.

\section{Similarity Test}

When the chosen XGBoost and NN models are selected, a test is done to ensure that their HE versions function correctly. A random subset of the test set is selected and inference on is performed using both the plaintext and HE models. The results from the HE model are decrypted, converted to binary, and compared with the results of the plaintext model. If they are not the same, the test is deemed a failure.

A subset is used rather than the full test set, as encrypting the full test set would take a lot of compute and time. An even balance of 75 fraudulent and 75 legitimate transactions are used in the test. This style of testing for HE inference is discussed in Section \ref{sec: further-he-testing}.

The HE XGBoost passed this test with all tested combinations of hyperparameters. The HE NN failed this test whenever more than one hidden layer was used. 

\section{Fraud Detection Performance}

\subsection{Metrics Used}

There is no consensus in the fraud detection literature on which metrics are best for evaluation. Using accuracy (percentage of transactions labelled correctly) is not practical due to the unbalanced data. Labelling every transaction as legitimate would result in very high accuracy. 

In these results, the threshold used for converting a probability prediction into a binary value is 0.5, i.e. any transaction given at least 50\% chance of being fraudulent is labelled as fraudulent. Different Hosts/Clients may choose this threshold to be lower or higher, depending on how sensitive they want their detector to be.

Threshold-free metrics evaluate models on their output before conversion to binary, and threshold-based metrics evaluate the binary decision. There is little consensus in the fraud detection literature on which metrics are best. In Reproducible Machine Learning for Credit Card Fraud Detection - Practical Handbook a combination of both is recommended \cite{leborgne2022fraud} . In these results, the AUC ROC score and the Average Precision score are the threshold-free metrics chosen, and the percentage of each type of transaction labelled correctly are chosen as the threshold-based evaluation.

\subsection{Models Chosen}

\begin{figure}[H]
      \centering
      \includegraphics[width=0.9\textwidth]{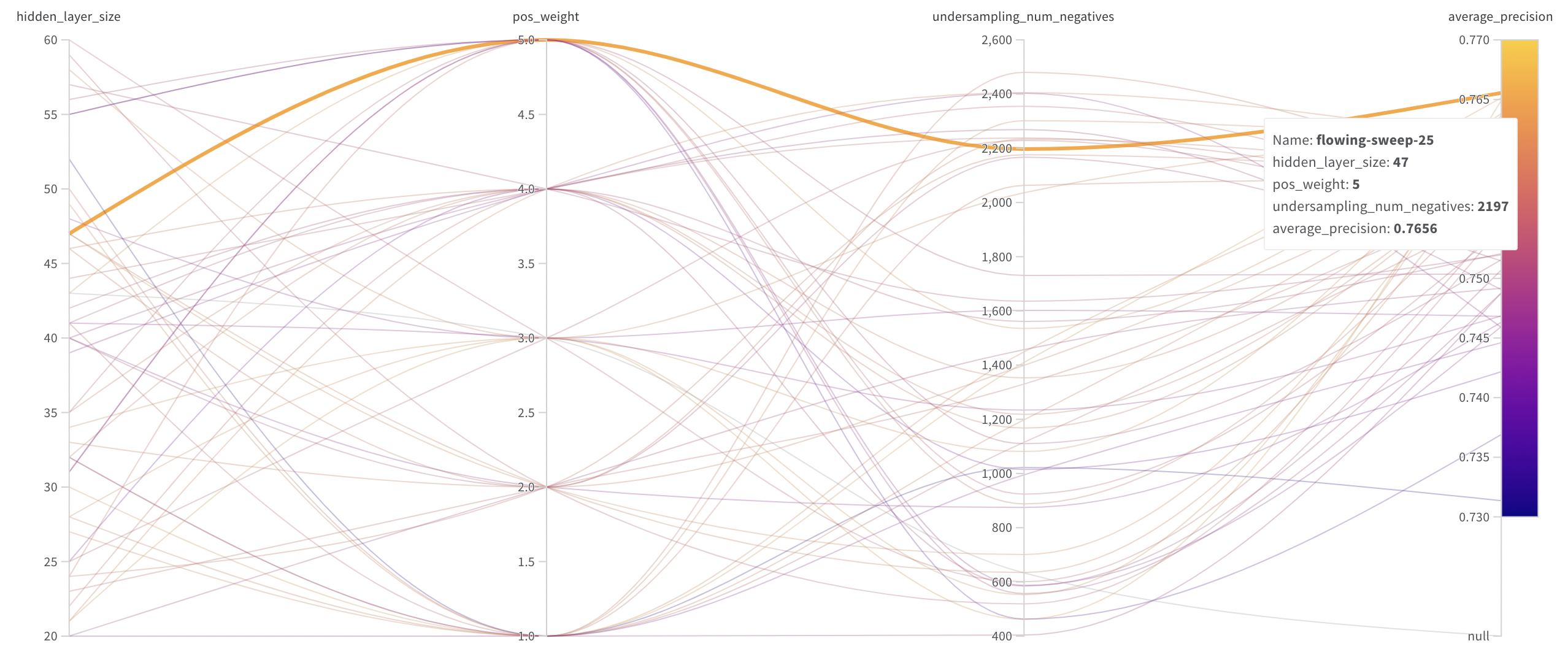}
      \caption{Selecting the NN model with the best Average Precision score on the ULB validation set.}
      \label{fig: ap-selection}
\end{figure}

The fraud detector to use on the test set is chosen based on how it performs on the validation set. Selecting for different metrics should provide different results on the test set. A comparison is made between the results obtained by the fraud detectors when each of the different metrics are selected for. Figure \ref{fig: ap-selection} shows selecting the NN model which had the best Average Precision score in a Sweep on the ULB dataset.

\begin{figure}[H]
      \centering
      \includegraphics[width=\textwidth]{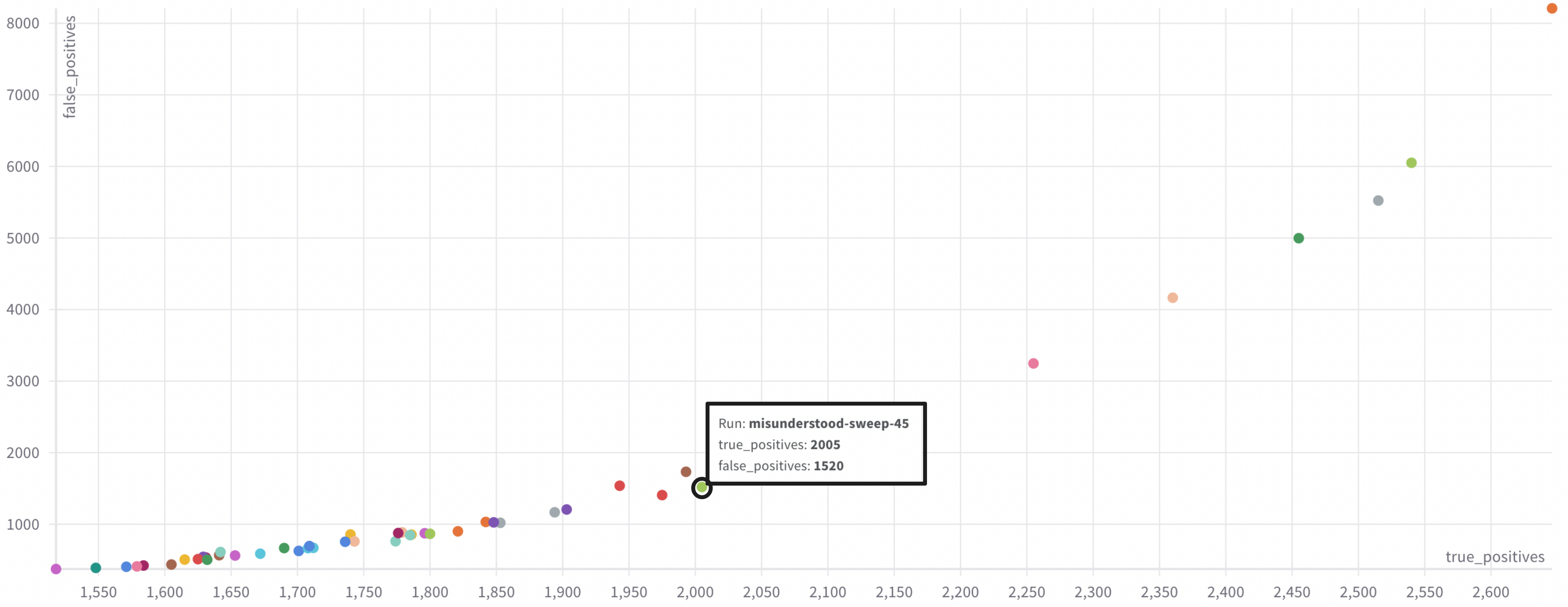}
      \caption{XGBoost model chosen by Inspection for Vesta dataset. Chart shows number of false negatives vs true positives detected on validation set.}
      \label{fig: inspection-selection}
\end{figure}

An extra model which was selected by inspection is also included in each case, to check if a person's judgement could result in better performance than by using a metric. The model which was deemed to have the best balance of correct labels on the validation set was selected in each case. An example of the XGBoost model selected by inspection on the Vesta dataset is shown in Figure \ref{fig: inspection-selection}.

The links to the relevant sets of training runs used for model selection, logged in W\&B, are listed in Table \ref{tab: training-run-links}.

\begin{table}[H]\centering
\caption{Links to training runs used for model selection}\label{tab: training-run-links}
\scriptsize
\begin{tabular}{lrrr}\toprule
\textbf{Model} &\textbf{Dataset} &\textbf{URL} \\\midrule
XGBoost &ULB &\url{https://wandb.ai/nuggimane/cc-ulb-xgboost/sweeps/o60jf75p} \\
&Vesta &\url{https://wandb.ai/nuggimane/cc-vesta-xgboost/sweeps/sff350wf} \\
& & \\
NN &ULB &\url{https://wandb.ai/nuggimane/cc-ulb-nn/sweeps/uyqqub3d}{} \\
&Vesta &\url{https://wandb.ai/nuggimane/cc-vesta-nn} \\
\bottomrule
\end{tabular}
\end{table}

\subsection{Detection Results}

The results for each dataset are shown in Tables \ref{tab: ulb-results} and \ref{tab: vesta-results}. Only models which passed the Similarity Test are used, i.e. HE NN models used have one hidden layer. The Fraudulent column shows the fraction of fraudulent transactions correctly detected as fraudulent, i.e. the recall. The Legitimate shows a similar fraction for the legitimate transactions. No NN models performed well on the Vesta dataset, so only one is shown.

\begin{table}[H]\centering
\caption{Detection results on ULB Dataset. Models with overall best performance are shown in bold.}\label{tab: ulb-results}
\scriptsize
\begin{tabular}{lrrrrrr}\toprule
\textbf{Model} &\textbf{Metric Selected For} &\textbf{AUC ROC} &\textbf{Average Precision} &\textbf{Fraudulent} &\textbf{Legitimate} \\\midrule
XGBoost &AUC ROC &0.989 &0.703 &0.88 &0.9689 \\
&\textbf{Average Precision} &\textbf{0.986} &\textbf{0.774} &\textbf{0.89} &\textbf{0.981} \\
&Inspection &0.982 &0.784 &0.81 &0.9977 \\
& & & & & \\
NN &AUC ROC &0.981 &0.744 &0.68 &0.9997 \\
&\textbf{Average Precision} &\textbf{0.981} &\textbf{0.739} &\textbf{0.77} &\textbf{0.9992} \\
&Inspection &0.982 &0.724 &0.77 &0.9993 \\
\bottomrule
\end{tabular}
\end{table}

\begin{table}[H]\centering
\caption{Detection results on Vesta Dataset. Models with overall best performance are shown in bold.}\label{tab: vesta-results}
\scriptsize
\begin{tabular}{lrrrrrr}\toprule
\textbf{Model} &\textbf{Metric Selected For} &\textbf{AUC ROC} &\textbf{Average Precision} &\textbf{Fraudulent} &\textbf{Legitimate} \\\midrule
XGBoost &AUC ROC &0.892 &0.498 &0.38 &0.9929 \\
&Average Precision &0.895 &0.508 &0.37 &0.9944 \\
&\textbf{Inspection} &\textbf{0.892} &\textbf{0.483} &\textbf{0.47} &\textbf{0.9789} \\
& & & & & \\
NN &\textbf{AUC ROC} &\textbf{0.725} &\textbf{0.142} &\textbf{0.72} &\textbf{0.5789} \\
\bottomrule
\end{tabular}
\end{table}

\section{Benchmarking}

Next, the results from the benchmarking of various latencies and storage sizes in the system are shown. As the NN model for the Vesta dataset failed to work as a fraud detector, it was excluded from further HE implementation benchmarking.

\subsection{Conditions for Benchmarking}

Latency benchmarking was run on an Intel Core i7-6700K CPU @ 4.00GHz  with an NVIDIA GeForce GTX1080 GPU. The models used for benchmarking were those with the overall best performance, as shown in Figures \ref{tab: ulb-results} and \ref{tab: vesta-results}.

The CKKS scheme's encryption context can be tuned to achieve better performance. Using higher poly modulus degree and coefficient modulus results in better precision, but also higher latency. The HE NN for the ULB dataset had 47 neurons in its layer, which allowed for a bit precision of 30 to pass the Similarity Test. 16384 was chosen as the poly modulus degree, with a coefficient modulus of [40, 30, 30, 30, 40].

The HE XGBoost used for benchmarking on ULB had a max depth of 9 and 51 trees, while the model used for on Vesta had a max depth of 7 and 92 trees. A maximum depth of 7 used for all XGBoost models considered for Vesta, as testing showed that using a higher depth resulted in model encryption times that were impractical (over 12 minutes). Using a fixed depth of 7 kept the model encryption time around 3 minutes.

\subsection{Latency Results}

The training time for the NN depends on the number of epochs chosen, while XGBoost automatically chooses when to stop training. The number of epochs trained on the ULB dataset was 100. Although it still showed poor detection performance, more epochs (300) were used to train on the Vesta dataset. Early stopping could potentially be added to the NN when a targeted validation loss is reached, however, training time is less critical than the other latencies.

Table \ref{tab: latency-results} displays the results for the various latencies involved in training and inference. For the sub-second metrics which require extra precision, 100 runs were done each time, selecting the fastest time found.

\begin{adjustwidth}{-2.5 cm}{-2.5cm}\centering\begin{threeparttable}[H]
\caption{Latency Results}\label{tab: latency-results}
\scriptsize
\begin{tabular}{lrrrrrrr}\toprule
\textbf{Model} &\textbf{Dataset} &\textbf{Training} &\textbf{Model Encryption} &\textbf{Transaction Encryption} &\textbf{Plaintext Inference} &\textbf{Encrypted Inference} \\\midrule
XGBoost &ULB &0.5s &13s &174ms &0.7ms &6ms \\
&Vesta &6s &2mins &2s &5.5ms &22ms \\
& & & & & & \\
NN &ULB &22s &N/A &10ms &0.08ms &296ms \\
\bottomrule
\end{tabular}
\end{threeparttable}\end{adjustwidth}

\subsection{Storage}

Table \ref{tab: storage-results} compares the sizes of transactions and models in their plaintext and ciphertext forms. These are important due to storage usage and loading times. The difference in the size of plaintext transactions between the XGBoost and NN models is because  they are stored as a Pandas DataFrame \cite{mckinney2011pandas} for XGBoost, whereas the transactions for the NN are stored as NumPy arrays \cite{harris2020array}.

\begin{table}[H]\centering
\caption{Storage Size Results}\label{tab: storage-results}
\scriptsize
\begin{tabular}{lrrrrrr}\toprule
\textbf{Model} &\textbf{Dataset} &\textbf{Plaintext Transaction} &\textbf{Encrypted Transaction} &\textbf{Plaintext Model} &\textbf{Encrypted Model} \\\midrule
XGBoost &ULB &1.2 kB &2.4 kB &92 kB &386 kB \\
&Vesta &10.7 kB &28.1 kB &896 kB &4512 kB \\
& & & & & \\
NN &ULB &0.387 kB &641 kB &8 kB &N/A \\
\bottomrule
\end{tabular}
\end{table}
\chapter{Discussion}
In this section the performance of the models are discussed, limitations of the system are explained, and potential further work is proposed.

\section{HE Model Comparison}

\subsection{Performance on ULB Dataset}

On the ULB dataset, the XGBoost model was successfully implemented as a CC fraud detector, with the best model correctly classifying 89\% of the fraudulent transactions and 98.1\% of legitimate transactions. The NN model also found success, with the best model achieving results of 77\% and 99.93\%. XGBoost was better at detecting fraudulent transactions, but the NN model produced less false positives.

In the latency benchmarking, HE XGBoost had an encrypted inference time of 6ms, while HE NN had a time of 296ms. When the encryption time of the transaction is added, the total latencies become 180ms and 306ms respectively. As the storage size of an encrypted HE XGBoost transaction is only 2.4kB, this would also be transmitted faster over the internet than the 641 kB transaction encrypted for HE NN.

Both models were successful on this dataset. HE XGBoost performs much faster, but the simpler deployment of HE NN may still make it the preferred choice.

\subsection{Performance on Vesta Dataset}

On the Vesta dataset, HE XGBoost found worse performance than on ULB, but the best model still correctly classified 47\% of fraudulent transactions and 97.89\% of legitimate transactions. The NN model failed to work as a fraud detector, likely due to the limited depth. Although it detected 72\% of the fraudulent transactions, it also labelled 42\% of legitimate transactions as fraudulent, which is unacceptable for a fraud detection system.

HE XGBoost had a fast inference time of 23ms, but saw a transaction encryption time of 2 seconds. Better encryption times could make HE XGBoost a decent model for this dataset.

\subsection{Metric Selected For}

No single basis for model choice between AUC ROC, Average Precision and Inspection consistently provided the best detection performance on the test set. For XGBoost on ULB, selecting for Average Precision produced the best results, while on the Vesta dataset, selecting by Inspection was the better option. The best metric for NN selection on ULB was probably Average Precision again, but very similar performance was found by Inspection. Overall, using Average Precision and the Inspection method proved to be the best option for these datasets.

\section{Limitations}

\subsection{Encrypted Outputs}
\label{sec: limitation-encrypted}

The models proposed above produce an encrypted output, which can only be decrypted using the secret key, which is held by the Client. This creates a limitation on the \hyperref[sec: use-cases-customer-bank]{Client \& Bank} use case, as the bank Host will be unable to determine if a given transaction is fraudulent, as the result is encrypted. Two possible solutions for this problem are as follows:

\begin{enumerate}
  \item The Client's secret key is used in a TEE to decrypt the output. This would involve Client encrypting their secret key using a key shared with Host and transmitting it to a TEE. Inside the TEE, the secret key is decrypted, and used to decrypt the output of the HE model. The secret key is discarded, and the Host receives whether the transaction was deemed fraudulent.
  
  \item The output of the model is not encrypted. This is possible using the XGBoost model, and other decision tree models. The inputs to the model are still encrypted, and still traverse the tree using OPE, but the leaf nodes are in plaintext, rather than using Paillier. This results in a model which takes encrypted inputs, and produces plaintext outputs.
  
  This does however introduce further security concerns. Host could theoretically derive information about an encrypted transaction by producing the same output of the model via brute-force, using transaction data generated by Host.
\end{enumerate}

One might ask why should HE be used at all if TEEs are going to be used in the pipeline anyways. The reason is that deploying complex attested code in TEEs is difficult to implement, especially at scale. Deploying a whole model inside a TEE could be very difficult, but using them for simple encryption/decryption steps to aid in the deployment of a HE model seems feasible.

\subsection{Lack of Real Data}

Using anonymized datasets means that our evaluation is undertaken on data that is not a direct representation of what would be used for training and inference in a production environment. Solutions to this would involve applying to gain access to private datasets, or using a simulator which accurately represents the distributions found in real data.

It is possible that some of the unknown features in these datasets could contain features engineered via aggregation, i.e. a value in a given transaction may contain information about previous transactions. This would undermine the \hyperref[sec: use-cases-customer-bank]{Client \& Bank} use case, which would need transaction features to be independent from other transactions.

\subsection{NN Model Depth}

The model depth of the NN model was limited to one hidden layer by the Similarity Test. Even when using the highest CKKS precision of 60 bits, deeper NNs still failed. As each hidden layer adds another square activation function, it is likely this is the issue.

This limitation means that if potential detection performance increases are available with increasing model size, that performance is unavailable to this implementation.

\subsection{Lack of Failure Assessment}

In a plaintext fraud detection system, production datapoints which result in false-positives or true-negatives can be reviewed, and changes to the model can be made accordingly. As the Host is unable to view the transaction data, they are unable to review why failures in the system occur.

One potential solution to this is that users could be asked to share information about a specific transaction if it was a transaction that caused a failure. If the transaction was not particularly private to the Client, they may be willing to give the plaintext version to improve the system. This is similar to how software applications may ask for the local system logs after a crash.

\subsection{HE XGBoost Information Leakage}

As the tree nodes of the model use OPE to make comparisons on the input data, if the plaintext model is known to the Host, they could watch the path taken by the input data down each of the trees. Knowing the result of each of the comparisons made allows Host to determine a range of possible values for each feature in the input data. This attack could be done by Host, or by a third-party attacker if they have access to the plaintext model, the encrypted model, and encrypted transaction data. A prevention for this attack could be to also run the training in the Secure Middle Server, as it was in the original algorithm proposed \cite{meng2020privacy}. But if Host has the data used for training and the information on how the model was trained, they could potentially make a replica model.

\section{Further Work}
\label{sec: further work}

\subsection{Deeper NN}
\label{sec: further-work-deeper-ckks}
To increase the depth of the CKKS NN, an interactive bootstrapping protocol could be introduced. Sending data back for decryption after every N layers in the model, but this would likely be too slow. It is also possible that the practical multiplicative depth could be increased by using other activation functions such as a polynomial approximation of ReLU.

\subsection{HE Inference speedups}
Parallelization across cores was implemented for the encryption of the PPXGB model, but it still struggled with long transaction encryption times. Further improvements could be reached by using GPU, FPGA, or ASIC.

\subsection{Implementation on Live Data}
As the datasets used have been anonymized, a true proof that this system is viable would be to train it on a private commercial dataset, and then test it on production transactions.

\subsection{Ensemble}
An ensemble of the two models could possibly be implemented, where the output of the HE XGBoost model and the HE NN model are combined to produce the final result. The leaf nodes of the XGBoost model could be encrypted using CKKS using the same key used for input to the NN model. This would allow the result of the two models to be added together. Alternatively, the two results could be returned to Client, and they would be combined after decryption.

\subsection{Feature Engineering using HE}
Feature engineering is widely used in CC fraud detection systems. This could potentially be implemented for encrypted transactions. An implementation would probably involve aggregating features across consecutive transactions via homomorphic addition, and producing mean values using homomorphic division.

\subsection{Testing of HE Models}
\label{sec: further-he-testing}
As HE models are slower than plaintext models, testing their performance on the full test set used for testing the plaintext model may be infeasible, as that also involves encrypting the full test set. Further work to build a confidence score of how well a HE model performs, or how similarly it performs to its plaintext equivalent could be valuable across domains.

\subsection{Blockchain Use Case}
Another proposed use case of the system relates to blockchain transaction fraud detection \cite{bhowmik2021comparative}. A transaction is made and added to a blockchain, along with extra transaction features. A smart contract \cite{zheng2020overview} then runs a HE fraud detector on these encrypted features. If the transaction is predicted as fraudulent, the funds are not transferred.
\chapter{Conclusion}
In this paper, a system for private credit card fraud detection using homomorphic encryption was proposed. Two models, XGBoost and a feedforward classifier neural network, were trained as fraud detectors on plaintext data. Models were chosen based on their performance on a validation set. Choosing models based on their average precision score and by inspection produced the best performance on the test set. They were then converted to models using homomorphic encryption, for inference on encrypted transactions.

A comparison of the models found that the XGBoost model had better detection performance and faster latency on the chosen datasets than the neural network model. However, the complexity of deploying the encrypted XGBoost model may result in the neural network being preferred in some cases.
\label{sec: conclusion}
\bibliographystyle{unsrtnat}
\bibliography{9-bibs/he,9-bibs/cc-fraud,9-bibs/nn,9-bibs/xgboost,9-bibs/other-methods}

\begin{thebibliography}{65}
\providecommand{\natexlab}[1]{#1}
\providecommand{\url}[1]{\texttt{#1}}
\expandafter\ifx\csname urlstyle\endcsname\relax
  \providecommand{\doi}[1]{doi: #1}\else
  \providecommand{\doi}{doi: \begingroup \urlstyle{rm}\Url}\fi

\bibitem[Micciancio(2010)]{micciancio2010first}
Daniele Micciancio.
\newblock A first glimpse of cryptography's holy grail.
\newblock \emph{Communications of the ACM}, 53\penalty0 (3):\penalty0 96--96,
  2010.

\bibitem[Prakaashini and Rajamohana(2021)]{prakaashini2021comprehensive}
S~Prakaashini and SP~Rajamohana.
\newblock Comprehensive report on homomorphic technique in healthcare domain.
\newblock In \emph{2021 5th International Conference on Computing Methodologies
  and Communication (ICCMC)}, pages 1448--1453. IEEE, 2021.

\bibitem[Han et~al.(2019)Han, Hong, Cheon, and Park]{han2019logistic}
Kyoohyung Han, Seungwan Hong, Jung~Hee Cheon, and Daejun Park.
\newblock Logistic regression on homomorphic encrypted data at scale.
\newblock In \emph{Proceedings of the AAAI Conference on Artificial
  Intelligence}, volume~33, pages 9466--9471, 2019.

\bibitem[ecb(2021)]{ecb2021}
Seventh report on card fraud.
\newblock \emph{European Central Bank}, 2021.

\bibitem[Leonard(1993)]{leonard1993detecting}
Kevin~J Leonard.
\newblock Detecting credit card fraud using expert systems.
\newblock \emph{Computers \& industrial engineering}, 25\penalty0
  (1-4):\penalty0 103--106, 1993.

\bibitem[Maes et~al.(2002)Maes, Tuyls, Vanschoenwinkel, and
  Manderick]{maes2002credit}
Sam Maes, Karl Tuyls, Bram Vanschoenwinkel, and Bernard Manderick.
\newblock Credit card fraud detection using bayesian and neural networks.
\newblock In \emph{Proceedings of the 1st international naiso congress on neuro
  fuzzy technologies}, volume~7, 2002.

\bibitem[Patidar et~al.(2011)Patidar, Sharma, et~al.]{patidar2011credit}
Raghavendra Patidar, Lokesh Sharma, et~al.
\newblock Credit card fraud detection using neural network.
\newblock \emph{International Journal of Soft Computing and Engineering
  (IJSCE)}, 1\penalty0 (32-38), 2011.

\bibitem[Yang et~al.(2019)Yang, Zhang, Ye, Li, and Xu]{yang2019ffd}
Wensi Yang, Yuhang Zhang, Kejiang Ye, Li~Li, and Cheng-Zhong Xu.
\newblock Ffd: A federated learning based method for credit card fraud
  detection.
\newblock In \emph{International conference on big data}, pages 18--32.
  Springer, 2019.

\bibitem[Abdi and Williams(2010)]{abdi2010principal}
Herv{\'e} Abdi and Lynne~J Williams.
\newblock Principal component analysis.
\newblock \emph{Wiley interdisciplinary reviews: computational statistics},
  2\penalty0 (4):\penalty0 433--459, 2010.

\bibitem[Ge et~al.(2020)Ge, Gu, Chang, and Cai]{ge2020credit}
Dingling Ge, Jianyang Gu, Shunyu Chang, and JingHui Cai.
\newblock Credit card fraud detection using lightgbm model.
\newblock In \emph{2020 International Conference on E-Commerce and Internet
  Technology (ECIT)}, pages 232--236. IEEE, 2020.

\bibitem[Fu et~al.(2016)Fu, Cheng, Tu, and Zhang]{fu2016credit}
Kang Fu, Dawei Cheng, Yi~Tu, and Liqing Zhang.
\newblock Credit card fraud detection using convolutional neural networks.
\newblock In \emph{International conference on neural information processing},
  pages 483--490. Springer, 2016.

\bibitem[Le~Borgne et~al.(2022)Le~Borgne, Siblini, Lebichot, and
  Bontempi]{leborgne2022fraud}
Yann-A{\"e}l Le~Borgne, Wissam Siblini, Bertrand Lebichot, and Gianluca
  Bontempi.
\newblock \emph{Reproducible Machine Learning for Credit Card Fraud Detection -
  Practical Handbook}.
\newblock Universit{\'e} Libre de Bruxelles, 2022.
\newblock URL
  \url{https://github.com/Fraud-Detection-Handbook/fraud-detection-handbook}.

\bibitem[Dal~Pozzolo et~al.(2014)Dal~Pozzolo, Caelen, Le~Borgne, Waterschoot,
  and Bontempi]{dal2014learned}
Andrea Dal~Pozzolo, Olivier Caelen, Yann-Ael Le~Borgne, Serge Waterschoot, and
  Gianluca Bontempi.
\newblock Learned lessons in credit card fraud detection from a practitioner
  perspective.
\newblock \emph{Expert systems with applications}, 41\penalty0 (10):\penalty0
  4915--4928, 2014.

\bibitem[Sahin and Duman(2010)]{sahin2010detecting}
Y~Sahin and Ekrem Duman.
\newblock Detecting credit card fraud by decision trees and support vector
  machines.
\newblock In \emph{World Congress on Engineering 2012. July 4-6, 2012. London,
  UK.}, volume 2188, pages 442--447. International Association of Engineers,
  2010.

\bibitem[Sahin and Duman(2011)]{sahin2011detecting}
Yusuf Sahin and Ekrem Duman.
\newblock Detecting credit card fraud by ann and logistic regression.
\newblock In \emph{2011 international symposium on innovations in intelligent
  systems and applications}, pages 315--319. IEEE, 2011.

\bibitem[Loh et~al.(2008)]{loh2008classification}
Wei-Yin Loh et~al.
\newblock Classification and regression tree methods.
\newblock \emph{Encyclopedia of statistics in quality and reliability},
  1:\penalty0 315--323, 2008.

\bibitem[Shen et~al.(2007)Shen, Tong, and Deng]{shen2007application}
Aihua Shen, Rencheng Tong, and Yaochen Deng.
\newblock Application of classification models on credit card fraud detection.
\newblock In \emph{2007 International conference on service systems and service
  management}, pages 1--4. IEEE, 2007.

\bibitem[Xuan et~al.(2018)Xuan, Liu, Li, Zheng, Wang, and
  Jiang]{xuan2018random}
Shiyang Xuan, Guanjun Liu, Zhenchuan Li, Lutao Zheng, Shuo Wang, and Changjun
  Jiang.
\newblock Random forest for credit card fraud detection.
\newblock In \emph{2018 IEEE 15th international conference on networking,
  sensing and control (ICNSC)}, pages 1--6. IEEE, 2018.

\bibitem[Friedman et~al.(2000)Friedman, Hastie, and
  Tibshirani]{friedman2000additive}
Jerome Friedman, Trevor Hastie, and Robert Tibshirani.
\newblock Additive logistic regression: a statistical view of boosting (with
  discussion and a rejoinder by the authors).
\newblock \emph{The annals of statistics}, 28\penalty0 (2):\penalty0 337--407,
  2000.

\bibitem[Priscilla and Prabha(2020)]{priscilla2020influence}
C~Victoria Priscilla and D~Padma Prabha.
\newblock Influence of optimizing xgboost to handle class imbalance in credit
  card fraud detection.
\newblock In \emph{2020 Third International Conference on Smart Systems and
  Inventive Technology (ICSSIT)}, pages 1309--1315. IEEE, 2020.

\bibitem[Xu et~al.(2015)Xu, Wang, Chen, and Li]{xu2015empirical}
Bing Xu, Naiyan Wang, Tianqi Chen, and Mu~Li.
\newblock Empirical evaluation of rectified activations in convolutional
  network.
\newblock \emph{arXiv preprint arXiv:1505.00853}, 2015.

\bibitem[Ghosh and Reilly(1994)]{ghosh1994credit}
Sushmito Ghosh and Douglas~L Reilly.
\newblock Credit card fraud detection with a neural-network.
\newblock In \emph{System Sciences, 1994. Proceedings of the Twenty-Seventh
  Hawaii International Conference on}, volume~3, pages 621--630. IEEE, 1994.

\bibitem[Aleskerov et~al.(1997)Aleskerov, Freisleben, and
  Rao]{aleskerov1997cardwatch}
Emin Aleskerov, Bernd Freisleben, and Bharat Rao.
\newblock Cardwatch: A neural network based database mining system for credit
  card fraud detection.
\newblock In \emph{Proceedings of the IEEE/IAFE 1997 computational intelligence
  for financial engineering (CIFEr)}, pages 220--226. IEEE, 1997.

\bibitem[Al-Shabi(2019)]{al2019credit}
MA~Al-Shabi.
\newblock Credit card fraud detection using autoencoder model in unbalanced
  datasets.
\newblock \emph{Journal of Advances in Mathematics and Computer Science},
  33\penalty0 (5):\penalty0 1--16, 2019.

\bibitem[Jurgovsky et~al.(2018)Jurgovsky, Granitzer, Ziegler, Calabretto,
  Portier, He-Guelton, and Caelen]{jurgovsky2018sequence}
Johannes Jurgovsky, Michael Granitzer, Konstantin Ziegler, Sylvie Calabretto,
  Pierre-Edouard Portier, Liyun He-Guelton, and Olivier Caelen.
\newblock Sequence classification for credit-card fraud detection.
\newblock \emph{Expert Systems with Applications}, 100:\penalty0 234--245,
  2018.

\bibitem[Rivest et~al.(1978)Rivest, Adleman, Dertouzos, et~al.]{rivest1978data}
Ronald~L Rivest, Len Adleman, Michael~L Dertouzos, et~al.
\newblock On data banks and privacy homomorphisms.
\newblock \emph{Foundations of secure computation}, 4\penalty0 (11):\penalty0
  169--180, 1978.

\bibitem[Gentry(2009)]{gentry2009fully}
Craig Gentry.
\newblock Fully homomorphic encryption using ideal lattices.
\newblock In \emph{Proceedings of the forty-first annual ACM symposium on
  Theory of computing}, pages 169--178, 2009.

\bibitem[Acar et~al.(2018)Acar, Aksu, Uluagac, and Conti]{acar2018survey}
Abbas Acar, Hidayet Aksu, A~Selcuk Uluagac, and Mauro Conti.
\newblock A survey on homomorphic encryption schemes: Theory and
  implementation.
\newblock \emph{ACM Computing Surveys (Csur)}, 51\penalty0 (4):\penalty0 1--35,
  2018.

\bibitem[Bost et~al.(2014)Bost, Popa, Tu, and Goldwasser]{bost2014machine}
Raphael Bost, Raluca~Ada Popa, Stephen Tu, and Shafi Goldwasser.
\newblock Machine learning classification over encrypted data.
\newblock \emph{Cryptology ePrint Archive}, 2014.

\bibitem[Goldwasser and Micali(2019)]{goldwasser2019probabilistic}
Shafi Goldwasser and Silvio Micali.
\newblock Probabilistic encryption \& how to play mental poker keeping secret
  all partial information.
\newblock In \emph{Providing sound foundations for cryptography: on the work of
  Shafi Goldwasser and Silvio Micali}, pages 173--201. 2019.

\bibitem[Halevi and Shoup(2014)]{halevi2014algorithms}
Shai Halevi and Victor Shoup.
\newblock Algorithms in helib.
\newblock In \emph{Annual Cryptology Conference}, pages 554--571. Springer,
  2014.

\bibitem[Paillier(1999)]{paillier1999public}
Pascal Paillier.
\newblock Public-key cryptosystems based on composite degree residuosity
  classes.
\newblock In \emph{International conference on the theory and applications of
  cryptographic techniques}, pages 223--238. Springer, 1999.

\bibitem[Khedr et~al.(2015)Khedr, Gulak, and Vaikuntanathan]{khedr2015shield}
Alhassan Khedr, Glenn Gulak, and Vinod Vaikuntanathan.
\newblock Shield: scalable homomorphic implementation of encrypted
  data-classifiers.
\newblock \emph{IEEE Transactions on Computers}, 65\penalty0 (9):\penalty0
  2848--2858, 2015.

\bibitem[Gentry et~al.(2013)Gentry, Sahai, and Waters]{gentry2013homomorphic}
Craig Gentry, Amit Sahai, and Brent Waters.
\newblock Homomorphic encryption from learning with errors:
  Conceptually-simpler, asymptotically-faster, attribute-based.
\newblock In \emph{Annual Cryptology Conference}, pages 75--92. Springer, 2013.

\bibitem[Meng and Feigenbaum(2020)]{meng2020privacy}
Xianrui Meng and Joan Feigenbaum.
\newblock Privacy-preserving xgboost inference.
\newblock \emph{arXiv preprint arXiv:2011.04789}, 2020.

\bibitem[Boldyreva et~al.(2009)Boldyreva, Chenette, Lee, and
  O’neill]{boldyreva2009order}
Alexandra Boldyreva, Nathan Chenette, Younho Lee, and Adam O’neill.
\newblock Order-preserving symmetric encryption.
\newblock In \emph{Annual International Conference on the Theory and
  Applications of Cryptographic Techniques}, pages 224--241. Springer, 2009.

\bibitem[Gilad-Bachrach et~al.(2016)Gilad-Bachrach, Dowlin, Laine, Lauter,
  Naehrig, and Wernsing]{gilad2016cryptonets}
Ran Gilad-Bachrach, Nathan Dowlin, Kim Laine, Kristin Lauter, Michael Naehrig,
  and John Wernsing.
\newblock Cryptonets: Applying neural networks to encrypted data with high
  throughput and accuracy.
\newblock In \emph{International conference on machine learning}, pages
  201--210. PMLR, 2016.

\bibitem[Bos et~al.(2013)Bos, Lauter, Loftus, and Naehrig]{bos2013improved}
Joppe~W Bos, Kristin Lauter, Jake Loftus, and Michael Naehrig.
\newblock Improved security for a ring-based fully homomorphic encryption
  scheme.
\newblock In \emph{IMA International Conference on Cryptography and Coding},
  pages 45--64. Springer, 2013.

\bibitem[Al~Badawi et~al.(2018)Al~Badawi, Chao, Lin, Fook~Mun, Jie~Sim,
  Meng~Tan, Nan, Aung, and Ramaseshan~Chandrasekhar]{al2018towards}
Ahmad Al~Badawi, Jin Chao, Jie Lin, Chan Fook~Mun, Jun Jie~Sim, Benjamin~Hong
  Meng~Tan, Xiao Nan, Khin Mi~Mi Aung, and Vijay Ramaseshan~Chandrasekhar.
\newblock Towards the alexnet moment for homomorphic encryption: Hcnn, thefirst
  homomorphic cnn on encrypted data with gpus.
\newblock \emph{arXiv e-prints}, pages arXiv--1811, 2018.

\bibitem[Brutzkus et~al.(2019)Brutzkus, Gilad-Bachrach, and
  Elisha]{brutzkus2019low}
Alon Brutzkus, Ran Gilad-Bachrach, and Oren Elisha.
\newblock Low latency privacy preserving inference.
\newblock In \emph{International Conference on Machine Learning}, pages
  812--821. PMLR, 2019.

\bibitem[Fan and Vercauteren(2012)]{fan2012somewhat}
Junfeng Fan and Frederik Vercauteren.
\newblock Somewhat practical fully homomorphic encryption.
\newblock \emph{Cryptology ePrint Archive}, 2012.

\bibitem[Boemer et~al.(2019)Boemer, Costache, Cammarota, and
  Wierzynski]{boemer2019ngraph}
Fabian Boemer, Anamaria Costache, Rosario Cammarota, and Casimir Wierzynski.
\newblock ngraph-he2: A high-throughput framework for neural network inference
  on encrypted data.
\newblock In \emph{Proceedings of the 7th ACM Workshop on Encrypted Computing
  \& Applied Homomorphic Cryptography}, pages 45--56, 2019.

\bibitem[Cheon et~al.(2017)Cheon, Kim, Kim, and Song]{cheon2017homomorphic}
Jung~Hee Cheon, Andrey Kim, Miran Kim, and Yongsoo Song.
\newblock Homomorphic encryption for arithmetic of approximate numbers.
\newblock In \emph{International Conference on the Theory and Application of
  Cryptology and Information Security}, pages 409--437. Springer, 2017.

\bibitem[Chillotti et~al.(2020)Chillotti, Gama, Georgieva, and
  Izabach{\`e}ne]{chillotti2020tfhe}
Ilaria Chillotti, Nicolas Gama, Mariya Georgieva, and Malika Izabach{\`e}ne.
\newblock Tfhe: fast fully homomorphic encryption over the torus.
\newblock \emph{Journal of Cryptology}, 33\penalty0 (1):\penalty0 34--91, 2020.

\bibitem[Clet et~al.(2021)Clet, Stan, and Zuber]{clet2021bfv}
Pierre-Emmanuel Clet, Oana Stan, and Martin Zuber.
\newblock Bfv, ckks, tfhe: Which one is the best for a secure neural network
  evaluation in the cloud?
\newblock In \emph{International Conference on Applied Cryptography and Network
  Security}, pages 279--300. Springer, 2021.

\bibitem[Chabanne et~al.(2017)Chabanne, De~Wargny, Milgram, Morel, and
  Prouff]{chabanne2017privacy}
Herv{\'e} Chabanne, Amaury De~Wargny, Jonathan Milgram, Constance Morel, and
  Emmanuel Prouff.
\newblock Privacy-preserving classification on deep neural network.
\newblock \emph{Cryptology ePrint Archive}, 2017.

\bibitem[Sabt et~al.(2015)Sabt, Achemlal, and Bouabdallah]{sabt2015trusted}
Mohamed Sabt, Mohammed Achemlal, and Abdelmadjid Bouabdallah.
\newblock Trusted execution environment: what it is, and what it is not.
\newblock In \emph{2015 IEEE Trustcom/BigDataSE/ISPA}, volume~1, pages 57--64.
  IEEE, 2015.

\bibitem[Goldreich(1998)]{goldreich1998secure}
Oded Goldreich.
\newblock Secure multi-party computation.
\newblock \emph{Manuscript. Preliminary version}, 78:\penalty0 110, 1998.

\bibitem[Canillas et~al.(2018)Canillas, Talbi, Bouchenak, Hasan, Brunie, and
  Sarrat]{canillas2018exploratory}
R{\'e}mi Canillas, Rania Talbi, Sara Bouchenak, Omar Hasan, Lionel Brunie, and
  Laurent Sarrat.
\newblock Exploratory study of privacy preserving fraud detection.
\newblock In \emph{Proceedings of the 19th International Middleware Conference
  Industry}, pages 25--31, 2018.

\bibitem[V{\'a}zquez-Saavedra et~al.(2021)V{\'a}zquez-Saavedra,
  Jim{\'e}nez-Balsa, Loureiro-Acu{\~n}a, Fern{\'a}ndez-Veiga, and
  Pedrouzo-Ulloa]{vazquez202154}
A~V{\'a}zquez-Saavedra, G~Jim{\'e}nez-Balsa, J~Loureiro-Acu{\~n}a,
  M~Fern{\'a}ndez-Veiga, and A~Pedrouzo-Ulloa.
\newblock 54 homomorphic svm inference for fraud detection.
\newblock 2021.

\bibitem[Biewald(2020)]{wandb}
Lukas Biewald.
\newblock Experiment tracking with weights and biases, 2020.
\newblock URL \url{https://www.wandb.com/}.
\newblock Software available from wandb.com.

\bibitem[Granger and P{\'e}rez(2021)]{granger2021jupyter}
Brian Granger and Fernando P{\'e}rez.
\newblock Jupyter: Thinking and storytelling with code and data.
\newblock \emph{Authorea Preprints}, 2021.

\bibitem[Chen and Guestrin(2016)]{chen2016xgboost}
Tianqi Chen and Carlos Guestrin.
\newblock Xgboost: A scalable tree boosting system.
\newblock In \emph{Proceedings of the 22nd acm sigkdd international conference
  on knowledge discovery and data mining}, pages 785--794, 2016.

\bibitem[Paszke et~al.(2019)Paszke, Gross, Massa, Lerer, Bradbury, Chanan,
  Killeen, Lin, Gimelshein, Antiga, Desmaison, Kopf, Yang, DeVito, Raison,
  Tejani, Chilamkurthy, Steiner, Fang, Bai, and Chintala]{NEURIPS2019_9015}
Adam Paszke, Sam Gross, Francisco Massa, Adam Lerer, James Bradbury, Gregory
  Chanan, Trevor Killeen, Zeming Lin, Natalia Gimelshein, Luca Antiga, Alban
  Desmaison, Andreas Kopf, Edward Yang, Zachary DeVito, Martin Raison, Alykhan
  Tejani, Sasank Chilamkurthy, Benoit Steiner, Lu~Fang, Junjie Bai, and Soumith
  Chintala.
\newblock Pytorch: An imperative style, high-performance deep learning library.
\newblock In H.~Wallach, H.~Larochelle, A.~Beygelzimer, F.~d\textquotesingle
  Alch\'{e}-Buc, E.~Fox, and R.~Garnett, editors, \emph{Advances in Neural
  Information Processing Systems 32}, pages 8024--8035. Curran Associates,
  Inc., 2019.
\newblock URL
  \url{http://papers.neurips.cc/paper/9015-pytorch-an-imperative-style-high-performance-deep-learning-library.pdf}.

\bibitem[Benaissa et~al.(2021)Benaissa, Retiat, Cebere, and
  Belfedhal]{benaissa2021tenseal}
Ayoub Benaissa, Bilal Retiat, Bogdan Cebere, and Alaa~Eddine Belfedhal.
\newblock Tenseal: A library for encrypted tensor operations using homomorphic
  encryption.
\newblock \emph{arXiv preprint arXiv:2104.03152}, 2021.

\bibitem[SEAL()]{sealcrypto}
SEAL.
\newblock {M}icrosoft {SEAL} (release 3.6).
\newblock \url{https://github.com/Microsoft/SEAL}, November 2020.
\newblock Microsoft Research, Redmond, WA.

\bibitem[Dal~Pozzolo et~al.(2015)Dal~Pozzolo, Caelen, Johnson, and
  Bontempi]{dal2015calibrating}
Andrea Dal~Pozzolo, Olivier Caelen, Reid~A Johnson, and Gianluca Bontempi.
\newblock Calibrating probability with undersampling for unbalanced
  classification.
\newblock In \emph{2015 IEEE Symposium Series on Computational Intelligence},
  pages 159--166. IEEE, 2015.

\bibitem[Guo and Berkhahn(2016)]{guo2016entity}
Cheng Guo and Felix Berkhahn.
\newblock Entity embeddings of categorical variables.
\newblock \emph{arXiv preprint arXiv:1604.06737}, 2016.

\bibitem[Ga{\v{z}}i et~al.(2014)Ga{\v{z}}i, Pietrzak, and
  Ryb{\'a}r]{gavzi2014exact}
Peter Ga{\v{z}}i, Krzysztof Pietrzak, and Michal Ryb{\'a}r.
\newblock The exact prf-security of nmac and hmac.
\newblock In \emph{Annual Cryptology Conference}, pages 113--130. Springer,
  2014.

\bibitem[Data61(2013)]{PythonPaillier}
CSIRO's Data61.
\newblock Python paillier library.
\newblock \url{https://github.com/data61/python-paillier}, 2013.

\bibitem[Albrecht et~al.(2018)Albrecht, Chase, Chen, Ding, Goldwasser,
  Gorbunov, Halevi, Hoffstein, Laine, Lauter, Lokam, Micciancio, Moody,
  Morrison, Sahai, and Vaikuntanathan]{HomomorphicEncryptionSecurityStandard}
Martin Albrecht, Melissa Chase, Hao Chen, Jintai Ding, Shafi Goldwasser, Sergey
  Gorbunov, Shai Halevi, Jeffrey Hoffstein, Kim Laine, Kristin Lauter, Satya
  Lokam, Daniele Micciancio, Dustin Moody, Travis Morrison, Amit Sahai, and
  Vinod Vaikuntanathan.
\newblock Homomorphic encryption security standard.
\newblock Technical report, HomomorphicEncryption.org, Toronto, Canada,
  November 2018.

\bibitem[McKinney et~al.(2011)]{mckinney2011pandas}
Wes McKinney et~al.
\newblock pandas: a foundational python library for data analysis and
  statistics.
\newblock \emph{Python for high performance and scientific computing},
  14\penalty0 (9):\penalty0 1--9, 2011.

\bibitem[Harris et~al.(2020)Harris, Millman, Van Der~Walt, Gommers, Virtanen,
  Cournapeau, Wieser, Taylor, Berg, Smith, et~al.]{harris2020array}
Charles~R Harris, K~Jarrod Millman, St{\'e}fan~J Van Der~Walt, Ralf Gommers,
  Pauli Virtanen, David Cournapeau, Eric Wieser, Julian Taylor, Sebastian Berg,
  Nathaniel~J Smith, et~al.
\newblock Array programming with numpy.
\newblock \emph{Nature}, 585\penalty0 (7825):\penalty0 357--362, 2020.

\bibitem[Bhowmik et~al.(2021)Bhowmik, Chandana, and
  Rudra]{bhowmik2021comparative}
Madhuparna Bhowmik, Tulasi Sai~Siri Chandana, and Bhawana Rudra.
\newblock Comparative study of machine learning algorithms for fraud detection
  in blockchain.
\newblock In \emph{2021 5th International Conference on Computing Methodologies
  and Communication (ICCMC)}, pages 539--541. IEEE, 2021.

\bibitem[Zheng et~al.(2020)Zheng, Xie, Dai, Chen, Chen, Weng, and
  Imran]{zheng2020overview}
Zibin Zheng, Shaoan Xie, Hong-Ning Dai, Weili Chen, Xiangping Chen, Jian Weng,
  and Muhammad Imran.
\newblock An overview on smart contracts: Challenges, advances and platforms.
\newblock \emph{Future Generation Computer Systems}, 105:\penalty0 475--491,
  2020.

\end{thebibliography}
\appendix
\renewcommand{\thechapter}{A\arabic{chapter}}
\chapter{Appendix}


\section{Plaintext-CKKS Neural Network Comparisons}

\begin{figure}[H]
      \centering
      \includegraphics[width=0.7\textwidth]{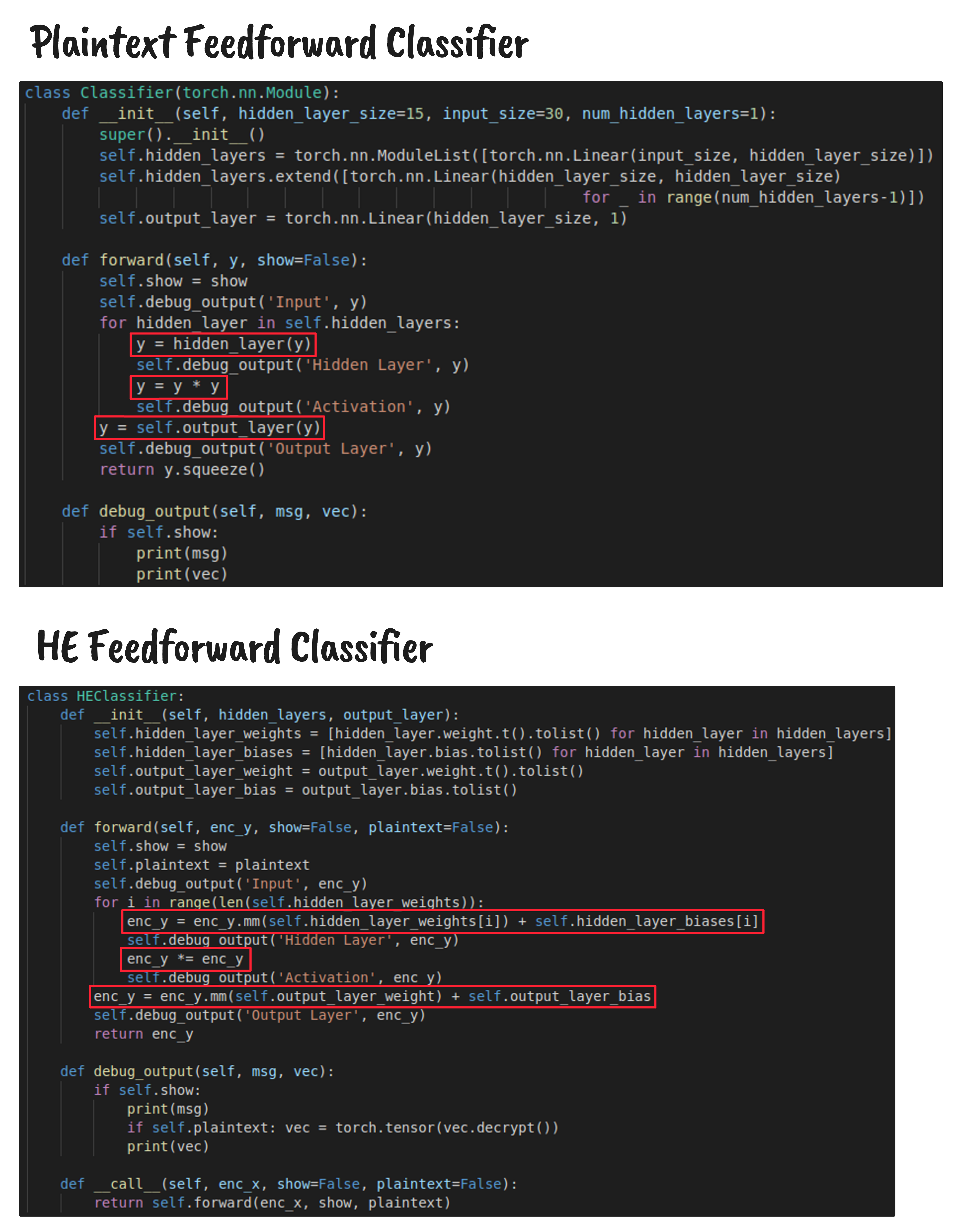}
      \caption{Comparison of code in plaintext and CKKS NNs. Corresponding layer operations are highlighted.}
      \label{fig: he-layers-comp}
\end{figure}

\begin{figure}[H]
      \centering
      \includegraphics[width=0.8\textwidth]{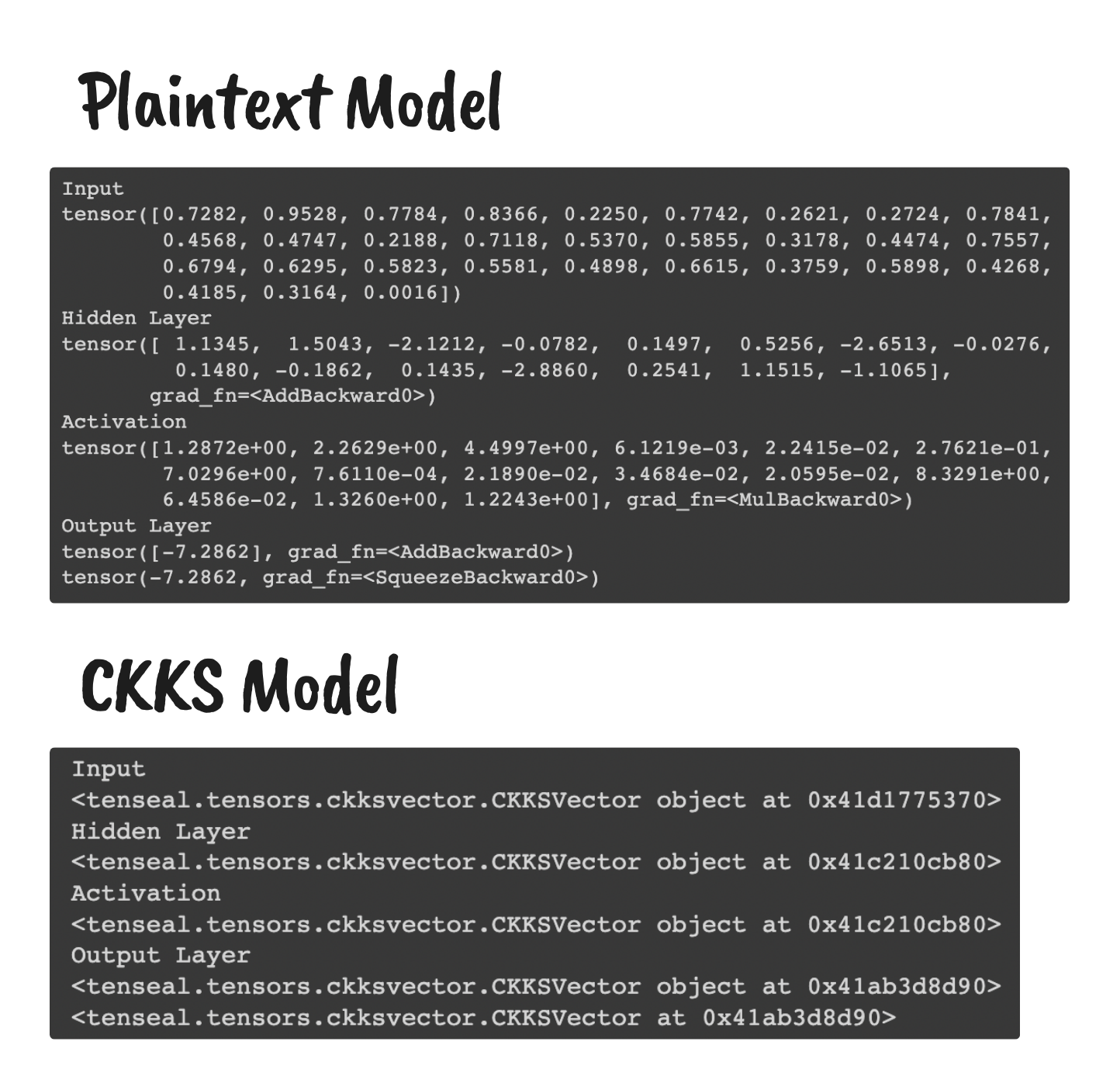}
      \caption{Comparison of data flow through plaintext and CKKS NNs.}
      \label{fig: nn-he-nn-comp}
\end{figure}

\section{Simulation}

\begin{figure}[H]
      \centering
      \includegraphics[width=0.8\textwidth]{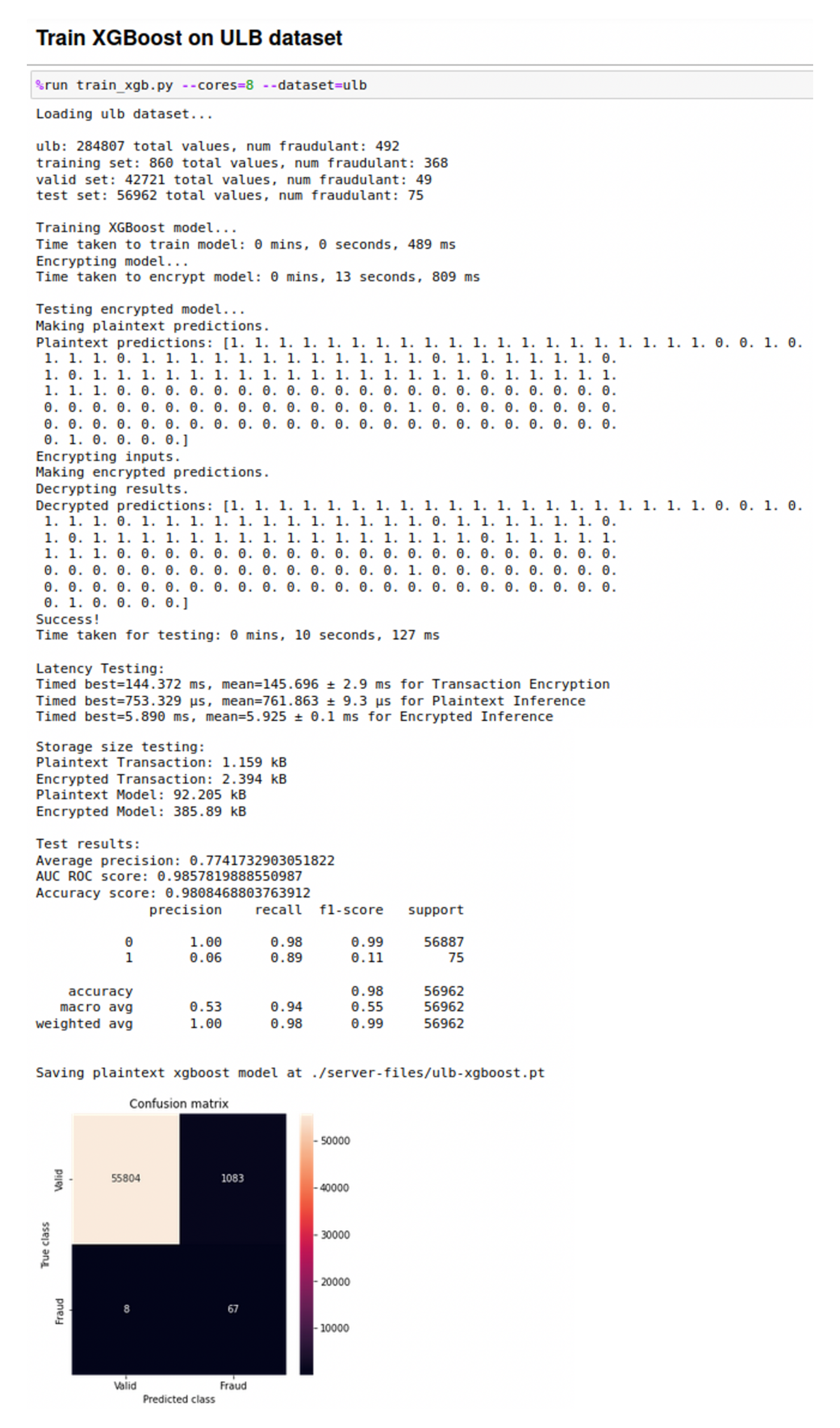}
      \caption{Training run of XGBoost on ULB dataset.}
      \label{fig: sim-ulb-training}
\end{figure}

\begin{figure}[H]
      \centering
      \includegraphics[width=\textwidth]{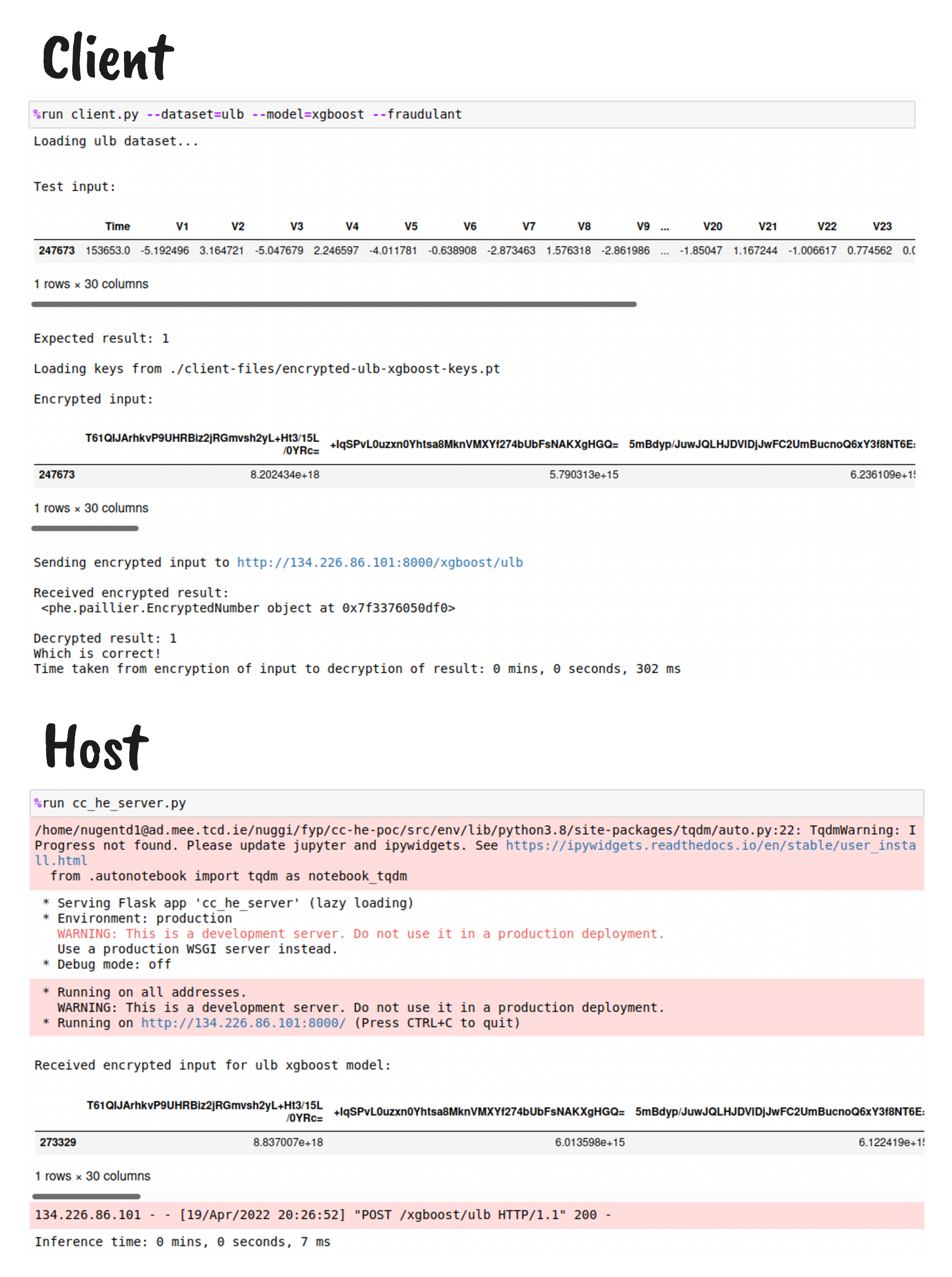}
      \caption{Interaction in Simulation between Client and Host}
      \label{fig: sim-client-host}
\end{figure}


\end{document}